\documentclass[preprint,superscriptaddress]{revtex4-2}

\usepackage{CJK}
\usepackage{times}
\usepackage{graphicx}
\usepackage{amsmath}
\usepackage{amssymb}
\usepackage[colorlinks,linkcolor=blue,citecolor=blue,urlcolor=blue,hyperindex,pdfstartview=FitH,plainpages=false]{hyperref}
\usepackage{float}
\usepackage{lineno}

\newcommand{\mJ}{mJ/cm$^2$}

\newcommand{\GT}{GdTe$_3$}
\newcommand{\A}{$\rm{\mathring{A}}^{-1}$}

\begin{document}


\begin{CJK*}{GBK}{}

\title{Identification of Metastable Lattice Distortion Free Charge Density Wave at Photoinduced Interface via TRARPES}
\author{Shaofeng Duan}
\affiliation{Beijing National Laboratory for Condensed Matter Physics, Institute of Physics, Chinese Academy of Sciences, Beijing 100190, China.}
\affiliation{Key Laboratory of Artificial Structures and Quantum Control (Ministry of Education), School of Physics and Astronomy, Shanghai Jiao Tong University, Shanghai 200240, China}
\author{Binshuo Zhang}
\affiliation{School of Physical Science and Technology, ShanghaiTech University, Shanghai 201210, China}
\author{Zihao Wang}
\author{Shichong Wang}
\author{Lingxiao Gu}
\author{Haoran Liu}
\author{Jiongyu Huang}
\author{Jianzhe Liu}
\affiliation{Key Laboratory of Artificial Structures and Quantum Control (Ministry of Education), School of Physics and Astronomy, Shanghai Jiao Tong University, Shanghai 200240, China}
\author{Dong Qian}
\affiliation{Key Laboratory of Artificial Structures and Quantum Control (Ministry of Education), School of Physics and Astronomy, Shanghai Jiao Tong University, Shanghai 200240, China}
\affiliation{Tsung-Dao Lee Institute, Shanghai Jiao Tong University, Shanghai 200240, China}
\affiliation{Collaborative Innovation Center of Advanced Microstructures, Nanjing University, Nanjing 210093, China}
\author{Yanfeng Guo}
\affiliation{School of Physical Science and Technology, ShanghaiTech University, Shanghai 201210, China}
\affiliation{ShanghaiTech Laboratory for Topological Physics, Shanghai 201210, China}
\author{Wentao Zhang}
\email{wentaozhang@iphy.ac.cn}
\affiliation{Beijing National Laboratory for Condensed Matter Physics, Institute of Physics, Chinese Academy of Sciences, Beijing 100190, China.}
\affiliation{Key Laboratory of Artificial Structures and Quantum Control (Ministry of Education), School of Physics and Astronomy, Shanghai Jiao Tong University, Shanghai 200240, China}
\date {\today}
\begin{abstract}

The interplay between different degrees of freedom governs the emergence of correlated electronic states in quantum materials, with charge density waves (CDW) often coexisting with other exotic phases. Under thermal equilibrium, traditional CDW states are consequentially accompanied by structural phase transitions.
In contrast, ultrafast photoexcitation allows access to exotic states where a single degree of freedom dominates in the time domain, enabling the study of underlying physics without interference.
Here, we report the realization of a long-lived metastable CDW state without lattice distortion at the photoinduced interfaces in \GT~using time- and angle-resolved photoemission spectroscopy.
After optical excitation above the CDW melting threshold, we identified emerged metastable interfaces through inverting the CDW-coupled lattice distortions, with lifetimes on the order of 10 picoseconds. These photoinduced interfaces represent a novel CDW state lacking the usual amplitude mode and lattice distortions, allowing quantification of the dominant role of electronic instabilities in CDW order.
This work provides a new approach to disentangling electronic instabilities from electron-phonon coupling using a nonequilibrium method.

\end{abstract}
\maketitle
\end{CJK*}

\section*{Introduction}
Understanding the mechanisms behind phase transitions and manipulating the emergence of exotic states in strongly correlated systems are central focuses in condensed matter physics. Charge density wave (CDW), characterized by periodic modulations in electron density, is a common feature in the phase diagrams of strongly correlated materials. It often coexists with or competes against unconventional superconductivity, as identified in cuprate high-temperature superconductors \cite{Keimer2015} and kagome systems \cite{Zheng2022}. Unraveling the driving force behind the CDW transition could provide key insights into the unconventional relationship between these neighboring phases. Owing to the coupled electronic and lattice degrees of freedom, the CDW transition is typically accompanied by periodic lattice distortions, raising the questions of whether the transition is driven primarily by electronic instability, electron-phonon coupling, or the distinct contributions of each \cite{Rossnagel2011b, Grner1994}. 
Numerous studies have shown that the CDW mechanism is material dependent, potentially arising from Fermi surface nesting in low-dimensional systems \cite{Whangbo1991, Brouet2008}, electron-phonon couplings in transition-metal dichalcogenides \cite{Varma1983, Xi2015}, electronic correlations in cuprates \cite{Fujita2014, Gerber2015}, or excitonic interactions in semimetals \cite{Cheng2022, Gao2024, Song2023}. Nonetheless, accurately quantifying the contributions of various interactions to the CDW transition remains a challenge for equilibrium experimental approaches. A promising strategy to tackle this issue could involve decoupling the electronic and lattice degrees of freedom, for instance, creating a pure CDW state without lattice distortion or a hidden state characterized solely by lattice distortions.

Ultrafast optical excitation has emerged as a promising knob for exploring and controlling the properties of matter in non-equilibrium conditions. This approach allows the isolation of various degrees of freedom in the time domain, offering new insights into their dynamics and interactions \cite{Bao2022, Torre2021, Xu2022, Perfetti2007}. Optical excitation can transiently quench the electronic order primarily through electron-electron scattering while preserving the low-temperature lattice symmetry on femtosecond timescales, as identified in excitonic insulators \cite{Katsumi2023, Tang2020}, CDW systems \cite{Rohwer2011, Maklar2021}, and iron-based superconductors \cite{Yang2022}. In addition, ultrafast laser pulses have been demonstrated to be a powerful tool to manipulate material properties, such as inducing exotic domains within them \cite{Kogar2020, Gerasimenko2019, Duan2021}. Through inhomogeneous excitation by infrared lasers, it is possible to reverse the CDW order within the material, creating domain boundaries that separate the initial phase from the inverted phase \cite{Yusupov2010, Trigo2021, Duan2021, kong2024, Duan2023}.  The interaction between opposing phases of adjacent domains may neutralize the periodic lattice distortions and restore the structural symmetry at these boundaries, resulting in the emergence of an exotic phase unattainable through thermal methods. However, the possibilities of realizing a pure CDW state without periodic lattice distortion at the domain boundary, providing a novel approach for studying the CDW transition, remain open and have not yet been experimentally explored.

In this article, we examine the quasi-two-dimensional CDW material GdTe$_3$ using high-resolution time- and angle-resolved photoemission spectroscopy (TRARPES). We show that ultrafast laser pulses melt the CDW order within hundreds of femtoseconds at a critical fluence of $F_\text{c}$ $\approx$ 0.34 \mJ, accompanied by the collapse of lattice distortions.
Furthermore, we identify the sectionally inverted CDW states above $F_\text{c}$, leading to macroscopic interfaces parallel to the sample surface on the picosecond timescales. The photoinduced interface is a metastable CDW state, characterized by the presence of a gap but lacking CDW-coupled lattice distortions. 
In addition, the energy gap at the interface is about 7 meV smaller than those at the nearby initial or inverted CDW states, suggesting that the structural factors play a limited role in the CDW transition. Our studies provide a unique perspective on understanding the contributions of the electronic and structural orders in the CDW transition.

\section*{Results}

\begin{figure}
\centering\includegraphics[width = 0.7\columnwidth] {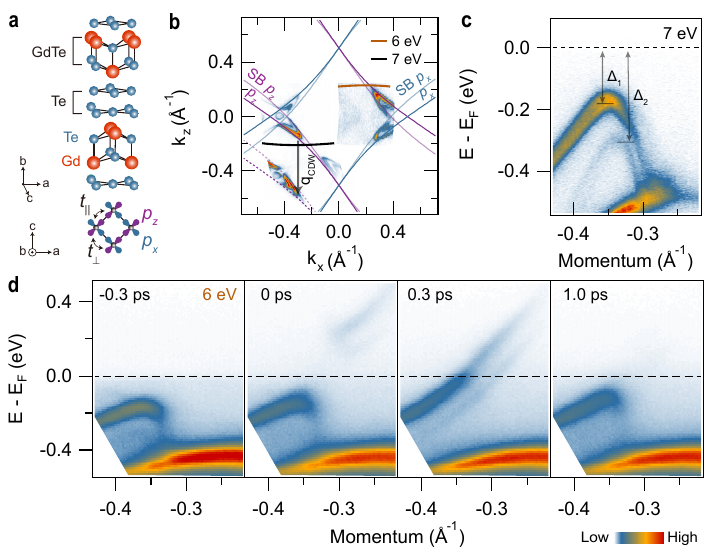}
\caption{
\textbf{Lattice and electronic structures of \GT.} \textbf{a,} Top: Schematic of the crystal structure. Bottom: Top view of the $p_x$ (green) and $p_z$ (purple) orbitals in the squared Te planes. The arrows indicate the orbital coupling along ($t{_\parallel}$) and perpendicular ($t_{\perp}$) to the chains. \textbf{b,} Fermi surface contours near the Fermi energy measured by 7-eV and 6-eV laser. The green and purple lines correspond to the $p_x$ and $p_z$ orbitals calculated with the tight-binding model, respectively. The black arrow indicates the CDW wavevector. Shadow bands (SB) and folded bands (FB) are noted. \textbf{c,} Equilibrium electronic structures at an equilibrium temperature of 4 K by 7-eV laser. \textbf{d,} Nonequilibrium electronic structures with a pump fluence of 0.64 \mJ~at selected delay times.} 
\label{Fig1}
\end{figure}

The material we investigated is the rare-earth tri-telluride system RTe$_3$ (R = Gd), and the driving force behind the CDW transition remains controversial, possibly driven by Fermi surface nesting \cite{Brouet2008, Brouet2004} or the enhanced electron-phonon coupling \cite{Johannes2008, Maschek2015, Eiter2013}. \GT~possess quasi-2D layered orthorhombic crystal structures, where GdTe slabs are sandwiched between two decoupled Te layers, forming a basic building block consisting of two Te square layers and double puckered GdTe slabs (Fig. \ref{Fig1}(a), top). A slight glide between the two Te planes along the c-axis results in a small in-plane anisotropy, leading to the preferred c-axis CDW and along with large lattice distortions \cite{Kim2006}. The electronic structures near the Fermi energy are dominated by the Te 5$p_x$ and 5$p_z$ orbitals (Fig. \ref{Fig1}a, bottom), and the Fermi surface contour, measured using 7-eV and 6-eV lasers, can be modeled by a tight-binding approach \cite{Brouet2008}. The Fermi surface is imperfectly nested, with some segments gapped and others remaining metallic (Fig. \ref{Fig1}b). Shadow and folded bands resulting from the CDW transition were clearly observed (Supplementary Note 1). The experimentally obtained CDW wave vector $q{_\text{CDW}}$ is about 0.43 \A~($\approx$ {2}/{7} $c^*$), consistent with previous reports \cite{Lee2023, Ru2008, Banerjee2013}.

The equilibrium electronic structure probed by 7-eV photons revealed two CDW gaps, approximately $\Delta_1$ = 175 meV and $\Delta_2$ = 290 meV (Fig. \ref{Fig1}c), consistent with results using 6-eV photons measured at the similar momentum cut shown in Fig. \ref{Fig1}d. The two bands with gap opening are a result of bilayer splitting \cite{Brouet2008, Liu2020}. Due to the absence of an a-axis CDW \cite{Banerjee2013}, the weaker spectral intensity around $k_x$ = 0 and $k_z$ = -0.4 \A ~in Fig. \ref{Fig1}b result from twinned domains with two perpendicular unidirectional CDWs \cite{Lee2023, Wang2022, Rettig2016}.
After optical excitation with a pump fluence of 0.64 \mJ, the unoccupied states of the CDW phase are clearly resolved at a delay time of 0 ps (Fig. \ref{Fig1}d). Subsequently, the CDW gap was transiently quenched at 0.3 ps, indicating a complete suppression of the CDW order. After 1 ps, the CDW gap was partially restored, and phase coherence recovered after several picoseconds \cite{GonzalezVallejo2022, Zong2019}.

\begin{figure*}
\centering\includegraphics[width = 1\columnwidth] {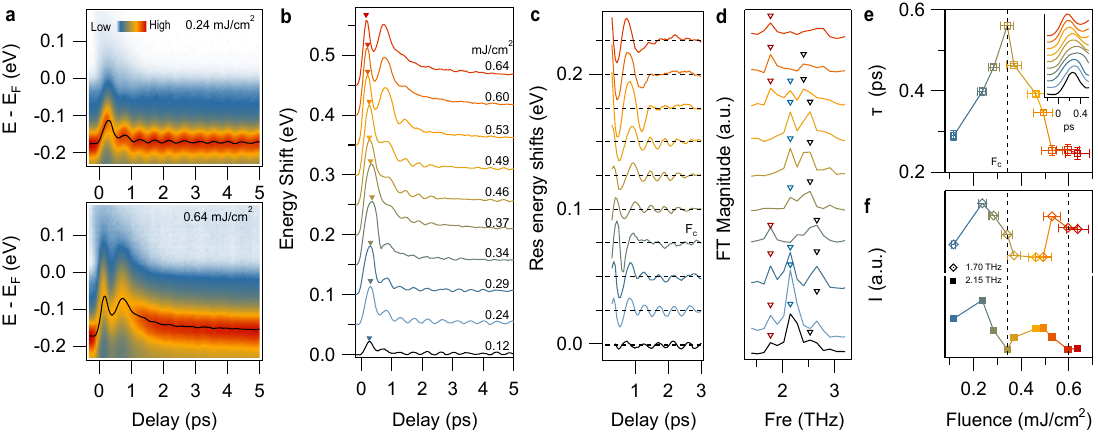}
\caption{
\textbf{Ultrafast electronic dynamics in \GT~at an equilibrium temperature of 4 K.} \textbf{a,} Time-dependent photoemission intensity at the valence band top as a function of binding energy with pump fluences of 0.24 \mJ~ (top) and 0.64 \mJ~ (bottom). The black solid lines represent the peak positions of the CDW bands determined from Lorentzian fitting. \textbf{b,} Time-dependent energy shifts of the CDW band at selected pump fluences. The solid inverted triangles indicate the time of maximum suppression of the CDW order. \textbf{c,} Residual energy shifts after subtracting smooth backgrounds from the curves shown in panel \textbf{b}. \textbf{d,} Fourier transform amplitudes for the coherent oscillation in panel \textbf{c}. \textbf{e,} The CDW suppression time as a function of the pump fluence. The inset shows the normalized energy shifts in panel b at delay times between 0 and 0.5 ps. \textbf{f,} The integrated intensity of the coherent phonon at frequencies of 1.77 and 2.15 THz as a function of the pump fluence. The black dashed lines represent the critical pump fluence for CDW order destruction.}
\label{Fig2}
\end{figure*}

To track the ultrafast evolution of the CDW after photoexcitation, we investigate the time-dependent photoemission intensity of the valence band top (Fig. \ref{Fig2}a). With a low pump fluence of 0.24 \mJ, the spectral intensity and binding energy of the CDW band exhibit pronounced oscillations up to 5 ps, resulting from optically-driven coherent phonons. In contrast, at a high pump fluence of 0.64 \mJ, which can fully quench the CDW order, the spectra only show anharmonic oscillations within 1 ps. This distinct behavior between high and low pump fluences is further evidenced by the photoinduced energy shifts as a function of delay time at various pump fluences (Fig. \ref{Fig2}b).
Notably, it takes the longest time of about 500 fs to reach the maximum suppression of the CDW gap at a critical fluence of $F_\text{c}$ $\approx$ 0.34 \mJ, leading to a $\lambda$-shaped curvature of the suppression time as a function of fluence (Fig. \ref{Fig2}e, Supplementary Note 2). This $\lambda$-shaped curvature is a signature of dynamical slowing down behavior near the critical point \cite{Zong2019a}.

It is intriguing that the photoinduced coherent oscillation of the energy shift is less pronounced or even disappears for delay times beyond 1 ps at the critical fluence of $F_\text{c}$ $\approx$ 0.34 \mJ~(Figs. \ref{Fig2}b and \ref{Fig2}c). 
This behavior is further supported by the Fourier transformation results, where three dominant components are observed with frequencies around 1.77, 2.15, and 2.65 THz below $F_\text{c}$, whereas the 2.15 THz component is absent precisely at $F_\text{c}$ and reappears above $F_\text{c}$ (Figs. \ref{Fig2}d and \ref{Fig2}f). The reappearance of the 2.15 THz phonon mode above $F_\text{c}$ suggests that the absence of oscillations at the critical fluence cannot be attributed to the overdamping effect, as the oscillation frequency softens gradually with increasing fluence \cite{GoncalvesFaria2024} and is eventually completely suppressed in the overdamped regime (Supplementary Note 3). Here, we attribute the most pronounced 2.15 THz mode to the amplitude mode \cite{Zong2019, Maklar2021, Leuenberger2015, Rettig2016, Schmitt2008, Yusupov2010, Rettig2014, Dutta, Lavagnini2010}, resulting from the periodic lattice distortions through the CDW transition.
The 2.65 THz frequency arises from other in-plane atomic motions and has been identified as a second amplitude mode in CeTe$_3$ \cite{Leuenberger2015}. Some studies associate the 1.77 THz mode with an optical phonon of the normal phase \cite{Yusupov2008, Moore2016}, while others consider it as another amplitude mode due to its strong temperature dependence \cite{Dutta, Lavagnini2010}. It should be noted that slightly above $F_\text{c}$, the 1.77 THz feature is also absent for reasons that remain unclear (Figs. \ref{Fig2}d and \ref{Fig2}f, top). The origins and complex interactions of these observed phonon modes are beyond the scope of this study.

The complete suppression of the amplitude mode at the CDW melting threshold $F_\text{c}$ indicates the diminishment of periodic lattice distortions, driving atomic motion towards its normal phase with a closed CDW gap within femtosecond timescales. The simultaneous collapse of the CDW and structural order complicates the identification of the driving force behind the transition in the time domain. Above the melting threshold, the reappearance of the amplitude mode strongly suggests light-induced CDW inversion \cite{Trigo2021, Duan2021}, and the further suppression of the amplitude mode at a critical fluence of about 0.59 \mJ~indicates the system finally recovers to its initial CDW state at long delay times (Fig. \ref{Fig2}f, bottom). As the light-induced inversions only take place above the critical fluence, the exponential reduction of the excitation fluence away from the sample surface could result in an interface between the inverted and original CDW states. The photoinduced interface is a novel state without lattice distortions and could persist for tens of picoseconds, as will be discussed later on.

\begin{figure}
\centering\includegraphics[width = 0.75\columnwidth] {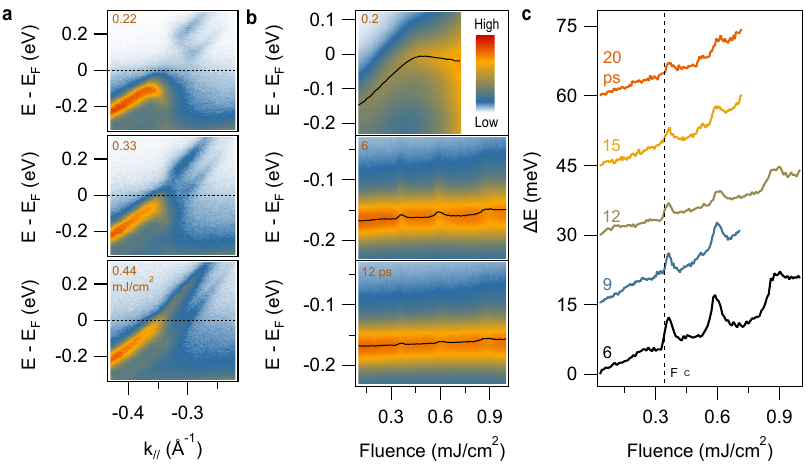}
\caption{
\textbf{Fluence-dependent TRARPES spectra in the CDW phase.} \textbf{a,} Nonequilibrium electronic structures at the delay time of 0.2 ps with fluences of 0.22, 0.33, and 0.44 \mJ. \textbf{b,} Fluence-dependent photoemission intensity at the momentum of -0.35 \A~with delay times of 0.2, 6, and 12 ps. The black lines indicate the peak positions of the CDW band. \textbf{c,} Fluence-dependent energy shifts of the CDW band at the momentum of -0.35 \A~with delay times of 6, 9, 12, 15, and 20 ps.}
\label{Fig3}
\end{figure}

The CDW melting threshold at $F_\text{c}$ is further supported by the fluence-dependent band shifts at a delay time of 0.2 ps, where the CDW gap collapses after photoexcitation due to transiently nonequilibrium quasiparticle population (Figs. \ref{Fig3}a and \ref{Fig3}b, top). As the fluence increases, the CDW gap (referring to the Fermi energy) saturates at about 0 meV at a delay time of 0.2 ps, and the notable photoemission intensity at highest fluence resulted from the pump-induced multi-photon effect (Fig. \ref{Fig3}b, top and Supplementary Note 4). The bending back for fluences above $F_\text{c}$ further confirms the distinct onset times for the CDW suppression (Fig. \ref{Fig2}e). Despite differing in magnitude below the Fermi energy, these two gaps of $\Delta_1$ and $\Delta_2$ due to bilayer splitting display a similar response to ultrafast optical excitation (Fig. \ref{Fig3}a and Supplementary Note 5), as they share the same origin driven by Fermi surface nesting. By finely tuning the fluence at a delay time of 6 ps, where the CDW gap has been already partially restored, distinct peak-dip features were observed in the energy shifts as a function of fluence (Figs. \ref{Fig3}b and \ref{Fig3}c). Below $F_\text{c}$, the energy shifts of the CDW band monotonically increased. Remarkably, the energy shifts exhibited a sharp peak feature at a fluence of approximately 0.36 \mJ, followed by broad dips for fluences between 0.42 and 0.56 \mJ. As the fluence was further increased, another maximum in energy shifts appeared at around 0.59 \mJ, followed by a decrease below 0.65 \mJ.
Since the energy shifts of the CDW band are governed by atomic displacements mediated by electron-phonon coupling, the transient matrix element effects have a negligible influence on the observed phenomena in our studies \cite{Boschini2020}.
These experimental findings are reminiscent of observations in the CDW material 1T-TiSe$_2$, where the peak-dip features in fluence-dependent energy shifts indicate the presence of phase-inversion-induced interface near the sample surface \cite{Duan2021}. Similar peak-dip features were also identified at delay times of 9, 12, 15, and 20 ps, though they became less pronounced at longer delay times (Fig. \ref{Fig3}c). The diminished peak-dip features at longer delay times suggest that out-of-plane phase coherence of the CDW order gradually restores over a timescale of 10 ps, leading to the annihilation of the photoinduced interface.

\begin{figure*}
\centering\includegraphics[width = 1\columnwidth] {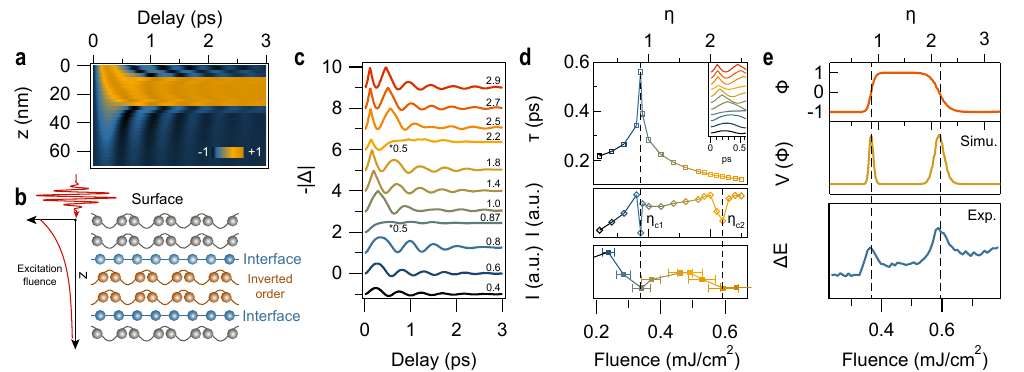}
\caption{
\textbf{Numerical solutions of the ultrafast dynamics of the order parameter based on the time- and spatially-dependent Ginzburg-Landau model.} \textbf{a,} Order parameter as a function of delay time and depth $z$ away from the sample surface at a fluence of $\eta = 4$. \textbf{b,} Schematic to show the phase-inversion induced interfacial CDW without lattice distortion. The solid lines indicate the amplitude of the atomic displacements. \textbf{c,} Order parameter of the surface ($z$ = 0) as a function of delay time for selected pump fluences $\eta$. \textbf{d,} Simulated fluence-dependent suppression time of the CDW order (top) and intensity of the amplitude mode (middle), along with experimental fluence-dependent integrated intensity of the amplitude mode at a frequency of 2.15 THz (bottom) from Fig. \ref{Fig2}\textbf{f}. The inset zooms in panel \textbf{c} between 0 and 0.6 ps. \textbf{e,} Fluence-dependent order parameter (top) and energy potential $V(\Phi)$ (middle) at a delay time of 6 ps, along with experimental fluence-dependent energy shifts (bottom) at the same delay time from Fig. \ref{Fig3}\textbf{c}. The x-axes in \textbf{d} and \textbf{e} are normalized to match the experimental and calculated data.}
\label{Fig4}
\end{figure*}

The formation of the photoinduced interface can be phenomenologically understood through a spatially- and time-dependent Ginzburg-Landau model with energy potential V($\Phi$, $z$, $t$) as a function of the order parameter $\Phi$ \cite{Duan2021, Trigo2021, Yusupov2010, kong2024}. The numerical solution of the order parameter as a function of delay time and depth ($z$) away from the surface at fluence $\eta = 4$ clearly reveals two sharp interfaces inside the sample, located between the initial ($\Phi = -1$) and inverted ($\Phi = +1$) states at delay times longer than 0.5 ps after photoexcitation (Fig. \ref{Fig4}a, Supplementary Note 6). The mechanism of such phase-inversion-induced interfaces is sketched in Fig. \ref{Fig4}b. The photoinduced interface gradually annihilates at longer delay times, possibly driven by the fluctuation of the order parameter \cite{kong2024}, which is not captured in the current phenomenological model. By finely tuning the pump fluences, phase-inversion-induced interfaces can be located near the sample surface at the simulated fluences of $\eta_\text{c1}$ = 0.87 and $\eta_\text{c2}$ = 2.15, and can be probed by a surface-sensitive technique such as ARPES in the current study. The simulated dynamics of the order parameters in Fig. \ref{Fig4}c qualitatively agree with the experimental results shown in Fig. \ref{Fig2}b, where the coherent oscillation disappears near the fluences for interfaces positioned at the sample surface.
Additionally, the simulated suppressed time of the CDW order also shows a prominent peak at the threshold $\eta_\text{c1}$, indicating the photoinduced transitions (Fig. \ref{Fig4}d, top). The magnitude of the CDW-coupled amplitude mode is also completely suppressed at the threshold $\eta_\text{c1}$, reemerges above $\eta_\text{c1}$ (Fig. \ref{Fig4}d, middle), and experiences further suppression at the threshold $\eta_\text{c2}$, aligning with the experimental observations (Fig. \ref{Fig4}d, bottom). 
The complete suppression of the CDW-coupled amplitude mode at excitation densities $\eta_\text{c1}$ and $\eta_\text{c2}$ is a signature of laser-induced order parameter inversion.

Furthermore, we delve into the ultrafast dynamics of the order parameters at the surface. For pump fluences below $\eta_\text{c1}$, the Landau potential undergoes slight modification after photoexcitation, initiating damped oscillations with the CDW amplitude mode that eventually recover to the initial state ($\Phi = -1$). As the pump fluence increases to $\eta_\text{c1} = 0.87$, the CDW order is fully quenched, and no coherent oscillations are observed at longer delay times (Fig. \ref{Fig4}c), where the free energy $V(\Phi,t)$ becomes a single-well function centered at $\Phi=0$. For pump fluences above $\eta_\text{c1}$, the CDW order is transiently quenched after photoexcitation, and the system transitions to its normal phase within about 1 ps, subsequently recovering to either its initial ($\Phi = -1$) or inverted ($\Phi = +1$) states at longer delay times depending on the incident excitation density.
The simulated order parameters as a function of pump fluence show that the free energy $V$ peaks at the photoinduced interfaces (Fig. \ref{Fig4}e, top and middle), qualitatively agreeing with the observation in the TRARPES experiments (Fig. \ref{Fig4}e, bottom). These simulations are consistent with the experiments, reproducing the ultrafast dynamics of CDW order after photoexcitation.

\section*{Discussion}

Comparing experimental observations with simulations, we find that the photoinduced interface in \GT~is remarkably sharp, nearly confined to the thickness of a single unit cell. For instance, in the fluence-dependent measurement at 6 ps, the peak width at 0.36 \mJ~is approximately 0.045 \mJ~(Fig.~\ref{Fig3}c). Given a penetration depth of 20 nm for the infrared pump laser \cite{Trigo2021}, the estimated depth corresponding to the absorption of the pump pulse energy within the peak range is about 2.67 nm---comparable to the lattice constant of 2.528 nm perpendicular to the cleaving surface in \GT. Similarly, the probing depth at a photon energy of 6 eV is approximately 3 nm \cite{Seah1979}, which closely matches both the lattice constant and the thickness of the domain interface, enabling effective investigation of the photoinduced interface state at the sample surface while minimizing bulk contributions. In addition, simulations have shown that the thickness and sharpness of the interface between initial and inverted states are primarily determined by the lattice constant perpendicular to the surface and the coherence length of the CDW order (Supplementary Note 7). A larger lattice constant and shorter coherence length would result in a thinner interface induced by the ultrafast laser. The opposing structural modulations on either side of the photoinduced interface significantly neutralize each other at the domain boundary, suppressing the periodic deformations and giving rise to an exotic metastable state free of lattice distortions. Experiments and calculations show that the observed interfaces at fluences of approximately 0.36 and 0.59 \mJ~are nearly confined with a unit cell, representing a nonequilibrium CDW state with gap opening (Figs. \ref{Fig3}b and \ref{Fig3}c) but no accompanying lattice distortions (Figs. \ref{Fig2}d and \ref{Fig2}f). The gap feature at the domain boundary is further supported by transient electronic structure measurements. The spectral difference between the domain boundary and the inverted CDW state reveals a further increment in spectral intensity within the CDW gap (Supplementary Note 8), indicating that the CDW gap at the photoinduced domain boundary is smaller than the bulk, even under lower pump fluence.

Furthermore, this lattice-distortion-free interface induced by ultrafast laser allows us to quantify the contributions of different degrees of freedom to the CDW transition. The pure CDW state without lattice distortions, located at the interfaces, serves as a hallmark of the electron-driven transition. Taking the Fermi-surface-nesting picture for an example, the electronic instability triggers CDW transition at a finite temperature, with the redistributed charge density driving atoms to new equilibrium positions, accompanied by periodic lattice distortions that further lower the overall energy \cite{Johannes2008, Clarke1994}. This suggests that the CDW states can persist even after suppressing lattice distortions in a nonequilibrium condition, as observed in the present study. In a structurally driven CDW transition, however, the modulated electronic density arises from the screening of the distorted lattice potential. In this scenario, suppressing the lattice distortions would simultaneously break down the CDW order \cite{Rossnagel2011b, Grner1994}. Thus, the identified metastable CDW state without lattice distortions at the interfaces indicates that the CDW transition in \GT~is driven by electronic interactions, with lattice distortion being a byproduct of the CDW transition. Furthermore, we find that the CDW gap at the interface is approximately 162 meV (7 meV smaller than that away from the interface, Fig. \ref{Fig3}c), with the estimated lattice contribution to the electronic gap around 48 meV, accounting for about 8\% of the total CDW gap. These results suggest that the structure plays a limited role in stabilizing the CDW state. 

Intriguingly, the photoinduced lattice distortion inversion is a common phenomenon in certain symmetry-broken systems, and can be evidenced by the emergence of anharmonic oscillations following the suppression of coherent phonon modes associated with the order parameter, as identified in CDW materials \cite{Trigo2019, Duan2021, Huber2014}, manganite \cite{Beaud2014}, and Peierls system \cite{Teitelbaum2018}. Such photoinduced phase inversion could potentially realize metastable interfaces without lattice distortion in other systems involving CDW order, including cuprates  \cite{Keimer2015} and kagome system \cite{Hu2023}, providing an effective way to clarify the origin of the exotic phase transitions through such nonequilibrium experiments. In addition, the photoinduced interfacial CDW develops at long delay times rather than immediately after photoexcitation, with the nonequilibrium electronic temperature of the system nearly recovered, accompanied by a reduced CDW gap at the photoinduced interface. In some superconductors, such as underdoped cuprates \cite{Keimer2015} and copper-intercalated 1T-TiSe$_2$ \cite{Morosan2006}, this photoinduced reduction of interfacial CDW order may enhance the superconductivity, as the CDW order potentially competes with superconducting pairing.

In conclusion, we identified a metastable CDW without lattice distortion at the photoinduced interfaces in the CDW material \GT, indicating the electronic origin of the CDW transition. These photoinduced interfaces form between the equilibrium and inverted phases. Leveraging this photoinduced pure CDW state, we quantitatively determined the contributions of intertwined electronic and structural degrees of freedom to the phase transition, finding that the electronic instability accounts for more than 90\% of the CDW gap. Experimentally, we observed that the lifetime of the photoinduced interfaces and the pure CDW state is on the order of 10 ps.
Our work presents a promising pathway for optically realizing exotic states that are inaccessible under thermal equilibrium, while also providing a novel approach for studying the mechanism behind correlated orders in quantum materials by isolating different degrees of freedom in the time domain.

\section*{methods}
\textbf{ARPES experiments.} The ultrafast electronic structure experiments were conducted using a TRARPES system with pump laser pulses at 700 nm (1.77 eV) and probe at 6 eV, operating at a repetition rate of 500 kHz \cite{Yang2019}. The steady-state thermal effect induced by the pump laser pulses is negligible for the pump fluences used in our time-resolved experiments (Supplementary Note 9). The time and energy resolutions were carefully set to approximately 113 fs and 16 meV, respectively, and a combined 7 eV laser was used to study the equilibrium electronic structure with an experimental energy resolution better than 0.5 meV \cite{Huang2022}. Sample position was dynamically corrected to ensure precise beam spot stability on the sample surface, achieving a precision better than 1-micron \cite{Duan2022a}. 

\textbf{Sample preparation.} High quality \GT~single crystal used in the experiments was synthesized via the self-flux method \cite{Liu2020}, and the sample was cleaved at 4 K under ultrahigh vacuum conditions of better than 3 $\times$ 10$^{-11}$ Torr.

\section*{DATA AVAILABILITY}
 The data that support the plots within this paper and other findings of this study are available
 from the corresponding author upon reasonable request. Correspondence and requests for materials should be addressed to W.T.Z. (wentaozhang@iphy.ac.cn).

\begin{acknowledgements}
W. T. Z. acknowledges support from the National Key R\&D Program of China (Grants No. 2021YFA1401800 and No. 2021YFA1400202) and National Natural Science Foundation of China (Grants No. 12141404) and the Natural Science Foundation of Shanghai (Grant Nos. 22ZR1479700 and 23XD1422200). S. F. D. acknowledges support from the China Postdoctoral Science Foundation (Grant No. 2022M722108) and China National Postdoctoral Program for Innovative Talents (Grant No. BX20230216) and National Natural Science Foundation of China (Grant No. 12304178).
Y. F. Guo acknowledges the National Key R\&D Program of China (Grant No. 2023YFA1406100) and the Double First-Class Initiative Fund of ShanghaiTech University.
\end{acknowledgements}

\section*{Author Contributions}

W.T.Z. proposed and designed the research. S.F.D., Z.H.W., L.X.G., S.C.W., H.R.L., J.Y.H., J.Z.L., and W.T.Z. contributed to the development and maintenance of the TRARPES system. S.F.D., Z.H.W., J.Y.H., and H.R.L. collected the TRARPES data. B.S.Z. and Y.F.G. prepared the single crystal sample. W.T.Z. wrote the paper with S.F.D. and all other authors. W.T.Z., S.F.D., D.Q., and all authors discussed the results and commented on the manuscript.

\section*{COMPETING INTERESTS}
The authors declare no competing interests.


\begin{thebibliography}{63}%
\makeatletter
\providecommand \@ifxundefined [1]{%
 \@ifx{#1\undefined}
}%
\providecommand \@ifnum [1]{%
 \ifnum #1\expandafter \@firstoftwo
 \else \expandafter \@secondoftwo
 \fi
}%
\providecommand \@ifx [1]{%
 \ifx #1\expandafter \@firstoftwo
 \else \expandafter \@secondoftwo
 \fi
}%
\providecommand \natexlab [1]{#1}%
\providecommand \enquote  [1]{``#1''}%
\providecommand \bibnamefont  [1]{#1}%
\providecommand \bibfnamefont [1]{#1}%
\providecommand \citenamefont [1]{#1}%
\providecommand \href@noop [0]{\@secondoftwo}%
\providecommand \href [0]{\begingroup \@sanitize@url \@href}%
\providecommand \@href[1]{\@@startlink{#1}\@@href}%
\providecommand \@@href[1]{\endgroup#1\@@endlink}%
\providecommand \@sanitize@url [0]{\catcode `\\12\catcode `\$12\catcode
  `\&12\catcode `\#12\catcode `\^12\catcode `\_12\catcode `\%12\relax}%
\providecommand \@@startlink[1]{}%
\providecommand \@@endlink[0]{}%
\providecommand \url  [0]{\begingroup\@sanitize@url \@url }%
\providecommand \@url [1]{\endgroup\@href {#1}{\urlprefix }}%
\providecommand \urlprefix  [0]{URL }%
\providecommand \Eprint [0]{\href }%
\providecommand \doibase [0]{https://doi.org/}%
\providecommand \selectlanguage [0]{\@gobble}%
\providecommand \bibinfo  [0]{\@secondoftwo}%
\providecommand \bibfield  [0]{\@secondoftwo}%
\providecommand \translation [1]{[#1]}%
\providecommand \BibitemOpen [0]{}%
\providecommand \bibitemStop [0]{}%
\providecommand \bibitemNoStop [0]{.\EOS\space}%
\providecommand \EOS [0]{\spacefactor3000\relax}%
\providecommand \BibitemShut  [1]{\csname bibitem#1\endcsname}%
\let\auto@bib@innerbib\@empty
\bibitem [{\citenamefont {Keimer}\ \emph {et~al.}(2015)\citenamefont {Keimer},
  \citenamefont {Kivelson}, \citenamefont {Norman}, \citenamefont {Uchida},\
  and\ \citenamefont {Zaanen}}]{Keimer2015}%
  \BibitemOpen
  \bibfield  {author} {\bibinfo {author} {\bibfnamefont {B.}~\bibnamefont
  {Keimer}}, \bibinfo {author} {\bibfnamefont {S.~A.}\ \bibnamefont
  {Kivelson}}, \bibinfo {author} {\bibfnamefont {M.~R.}\ \bibnamefont
  {Norman}}, \bibinfo {author} {\bibfnamefont {S.}~\bibnamefont {Uchida}},\
  and\ \bibinfo {author} {\bibfnamefont {J.}~\bibnamefont {Zaanen}},\
  }\bibfield  {title} {\bibinfo {title} {From quantum matter to
  high-temperature superconductivity in copper oxides},\ }\href
  {https://doi.org/10.1038/nature14165} {\bibfield  {journal} {\bibinfo
  {journal} {Nature}\ }\textbf {\bibinfo {volume} {518}},\ \bibinfo {pages}
  {179} (\bibinfo {year} {2015})}\BibitemShut {NoStop}%
\bibitem [{\citenamefont {Zheng}\ \emph {et~al.}(2022)\citenamefont {Zheng},
  \citenamefont {Wu}, \citenamefont {Yang}, \citenamefont {Nie}, \citenamefont
  {Shan}, \citenamefont {Sun}, \citenamefont {Song}, \citenamefont {Yu},
  \citenamefont {Li}, \citenamefont {Zhao}, \citenamefont {Li}, \citenamefont
  {Kang}, \citenamefont {Zhou}, \citenamefont {Liu}, \citenamefont {Xiang},
  \citenamefont {Ying}, \citenamefont {Wang}, \citenamefont {Wu},\ and\
  \citenamefont {Chen}}]{Zheng2022}%
  \BibitemOpen
  \bibfield  {author} {\bibinfo {author} {\bibfnamefont {L.}~\bibnamefont
  {Zheng}}, \bibinfo {author} {\bibfnamefont {Z.}~\bibnamefont {Wu}}, \bibinfo
  {author} {\bibfnamefont {Y.}~\bibnamefont {Yang}}, \bibinfo {author}
  {\bibfnamefont {L.}~\bibnamefont {Nie}}, \bibinfo {author} {\bibfnamefont
  {M.}~\bibnamefont {Shan}}, \bibinfo {author} {\bibfnamefont {K.}~\bibnamefont
  {Sun}}, \bibinfo {author} {\bibfnamefont {D.}~\bibnamefont {Song}}, \bibinfo
  {author} {\bibfnamefont {F.}~\bibnamefont {Yu}}, \bibinfo {author}
  {\bibfnamefont {J.}~\bibnamefont {Li}}, \bibinfo {author} {\bibfnamefont
  {D.}~\bibnamefont {Zhao}}, \bibinfo {author} {\bibfnamefont {S.}~\bibnamefont
  {Li}}, \bibinfo {author} {\bibfnamefont {B.}~\bibnamefont {Kang}}, \bibinfo
  {author} {\bibfnamefont {Y.}~\bibnamefont {Zhou}}, \bibinfo {author}
  {\bibfnamefont {K.}~\bibnamefont {Liu}}, \bibinfo {author} {\bibfnamefont
  {Z.}~\bibnamefont {Xiang}}, \bibinfo {author} {\bibfnamefont
  {J.}~\bibnamefont {Ying}}, \bibinfo {author} {\bibfnamefont {Z.}~\bibnamefont
  {Wang}}, \bibinfo {author} {\bibfnamefont {T.}~\bibnamefont {Wu}},\ and\
  \bibinfo {author} {\bibfnamefont {X.}~\bibnamefont {Chen}},\ }\bibfield
  {title} {\bibinfo {title} {Emergent charge order in pressurized kagome
  superconductor {CsV$_3$Sb$_5$}},\ }\href
  {https://doi.org/10.1038/s41586-022-05351-3} {\bibfield  {journal} {\bibinfo
  {journal} {Nature}\ }\textbf {\bibinfo {volume} {611}},\ \bibinfo {pages}
  {682} (\bibinfo {year} {2022})}\BibitemShut {NoStop}%
\bibitem [{\citenamefont {Rossnagel}(2011)}]{Rossnagel2011b}%
  \BibitemOpen
  \bibfield  {author} {\bibinfo {author} {\bibfnamefont {K.}~\bibnamefont
  {Rossnagel}},\ }\bibfield  {title} {\bibinfo {title} {On the origin of
  charge-density waves in select layered transition-metal dichalcogenides},\
  }\href {https://doi.org/10.1088/0953-8984/23/21/213001} {\bibfield  {journal}
  {\bibinfo  {journal} {J. Phys. Condens. Matter}\ }\textbf {\bibinfo {volume}
  {23}},\ \bibinfo {pages} {213001} (\bibinfo {year} {2011})}\BibitemShut
  {NoStop}%
\bibitem [{\citenamefont {Gr{\"u}ner}(1994)}]{Grner1994}%
  \BibitemOpen
  \bibfield  {author} {\bibinfo {author} {\bibfnamefont {G.}~\bibnamefont
  {Gr{\"u}ner}},\ }in\ \href
  {https://api.semanticscholar.org/CorpusID:116898332} {\emph {\bibinfo
  {booktitle} {Density Waves In Solids}}}\ (\bibinfo {year} {1994})\BibitemShut
  {NoStop}%
\bibitem [{\citenamefont {Whangbo}\ \emph {et~al.}(1991)\citenamefont
  {Whangbo}, \citenamefont {Canadell}, \citenamefont {Foury},\ and\
  \citenamefont {Pouget}}]{Whangbo1991}%
  \BibitemOpen
  \bibfield  {author} {\bibinfo {author} {\bibfnamefont {M.-H.}\ \bibnamefont
  {Whangbo}}, \bibinfo {author} {\bibfnamefont {E.}~\bibnamefont {Canadell}},
  \bibinfo {author} {\bibfnamefont {P.}~\bibnamefont {Foury}},\ and\ \bibinfo
  {author} {\bibfnamefont {J.~P.}\ \bibnamefont {Pouget}},\ }\bibfield  {title}
  {\bibinfo {title} {Hidden fermi surface nesting and charge density wave
  instability in low-dimensional metals},\ }\href
  {https://doi.org/10.1126/science.252.5002.96} {\bibfield  {journal} {\bibinfo
   {journal} {Science}\ }\textbf {\bibinfo {volume} {252}},\ \bibinfo {pages}
  {96} (\bibinfo {year} {1991})}\BibitemShut {NoStop}%
\bibitem [{\citenamefont {Brouet}\ \emph {et~al.}(2008)\citenamefont {Brouet},
  \citenamefont {Yang}, \citenamefont {Zhou}, \citenamefont {Hussain},
  \citenamefont {Moore}, \citenamefont {He}, \citenamefont {Lu}, \citenamefont
  {Shen}, \citenamefont {Laverock}, \citenamefont {Dugdale}, \citenamefont
  {Ru},\ and\ \citenamefont {Fisher}}]{Brouet2008}%
  \BibitemOpen
  \bibfield  {author} {\bibinfo {author} {\bibfnamefont {V.}~\bibnamefont
  {Brouet}}, \bibinfo {author} {\bibfnamefont {W.~L.}\ \bibnamefont {Yang}},
  \bibinfo {author} {\bibfnamefont {X.~J.}\ \bibnamefont {Zhou}}, \bibinfo
  {author} {\bibfnamefont {Z.}~\bibnamefont {Hussain}}, \bibinfo {author}
  {\bibfnamefont {R.~G.}\ \bibnamefont {Moore}}, \bibinfo {author}
  {\bibfnamefont {R.}~\bibnamefont {He}}, \bibinfo {author} {\bibfnamefont
  {D.~H.}\ \bibnamefont {Lu}}, \bibinfo {author} {\bibfnamefont {Z.~X.}\
  \bibnamefont {Shen}}, \bibinfo {author} {\bibfnamefont {J.}~\bibnamefont
  {Laverock}}, \bibinfo {author} {\bibfnamefont {S.~B.}\ \bibnamefont
  {Dugdale}}, \bibinfo {author} {\bibfnamefont {N.}~\bibnamefont {Ru}},\ and\
  \bibinfo {author} {\bibfnamefont {I.~R.}\ \bibnamefont {Fisher}},\ }\bibfield
   {title} {\bibinfo {title} {Angle-resolved photoemission study of the
  evolution of band structure and charge density wave properties in
  {$R{\text{Te}}_{3}$ ($R=\text{Y}$, La, Ce, Sm, Gd, Tb, and Dy)}},\ }\href
  {https://doi.org/10.1103/PhysRevB.77.235104} {\bibfield  {journal} {\bibinfo
  {journal} {Phys. Rev. B}\ }\textbf {\bibinfo {volume} {77}},\ \bibinfo
  {pages} {235104} (\bibinfo {year} {2008})}\BibitemShut {NoStop}%
\bibitem [{\citenamefont {Varma}\ and\ \citenamefont
  {Simons}(1983)}]{Varma1983}%
  \BibitemOpen
  \bibfield  {author} {\bibinfo {author} {\bibfnamefont {C.~M.}\ \bibnamefont
  {Varma}}\ and\ \bibinfo {author} {\bibfnamefont {A.~L.}\ \bibnamefont
  {Simons}},\ }\bibfield  {title} {\bibinfo {title} {Strong-coupling theory of
  charge-density-wave transitions},\ }\href
  {https://doi.org/10.1103/PhysRevLett.51.138} {\bibfield  {journal} {\bibinfo
  {journal} {Phys. Rev. Lett.}\ }\textbf {\bibinfo {volume} {51}},\ \bibinfo
  {pages} {138} (\bibinfo {year} {1983})}\BibitemShut {NoStop}%
\bibitem [{\citenamefont {Xi}\ \emph {et~al.}(2015)\citenamefont {Xi},
  \citenamefont {Zhao}, \citenamefont {Wang}, \citenamefont {Berger},
  \citenamefont {Forr\'o}, \citenamefont {Shan},\ and\ \citenamefont
  {Mak}}]{Xi2015}%
  \BibitemOpen
  \bibfield  {author} {\bibinfo {author} {\bibfnamefont {X.}~\bibnamefont
  {Xi}}, \bibinfo {author} {\bibfnamefont {L.}~\bibnamefont {Zhao}}, \bibinfo
  {author} {\bibfnamefont {Z.}~\bibnamefont {Wang}}, \bibinfo {author}
  {\bibfnamefont {H.}~\bibnamefont {Berger}}, \bibinfo {author} {\bibfnamefont
  {L.}~\bibnamefont {Forr\'o}}, \bibinfo {author} {\bibfnamefont
  {J.}~\bibnamefont {Shan}},\ and\ \bibinfo {author} {\bibfnamefont {K.~F.}\
  \bibnamefont {Mak}},\ }\bibfield  {title} {\bibinfo {title} {Strongly
  enhanced charge-density-wave order in monolayer {NbSe$_2$}},\ }\href
  {https://doi.org/10.1038/nnano.2015.143} {\bibfield  {journal} {\bibinfo
  {journal} {Nat. Nanotechnol.}\ }\textbf {\bibinfo {volume} {10}},\ \bibinfo
  {pages} {765} (\bibinfo {year} {2015})}\BibitemShut {NoStop}%
\bibitem [{\citenamefont {Fujita}\ \emph {et~al.}(2014)\citenamefont {Fujita},
  \citenamefont {Hamidian}, \citenamefont {Edkins}, \citenamefont {Kim},
  \citenamefont {Kohsaka}, \citenamefont {Azuma}, \citenamefont {Takano},
  \citenamefont {Takagi}, \citenamefont {Eisaki}, \citenamefont {ichi Uchida},
  \citenamefont {Allais}, \citenamefont {Lawler}, \citenamefont {Kim},
  \citenamefont {Sachdev},\ and\ \citenamefont {Davis}}]{Fujita2014}%
  \BibitemOpen
  \bibfield  {author} {\bibinfo {author} {\bibfnamefont {K.}~\bibnamefont
  {Fujita}}, \bibinfo {author} {\bibfnamefont {M.~H.}\ \bibnamefont
  {Hamidian}}, \bibinfo {author} {\bibfnamefont {S.~D.}\ \bibnamefont
  {Edkins}}, \bibinfo {author} {\bibfnamefont {C.~K.}\ \bibnamefont {Kim}},
  \bibinfo {author} {\bibfnamefont {Y.}~\bibnamefont {Kohsaka}}, \bibinfo
  {author} {\bibfnamefont {M.}~\bibnamefont {Azuma}}, \bibinfo {author}
  {\bibfnamefont {M.}~\bibnamefont {Takano}}, \bibinfo {author} {\bibfnamefont
  {H.}~\bibnamefont {Takagi}}, \bibinfo {author} {\bibfnamefont
  {H.}~\bibnamefont {Eisaki}}, \bibinfo {author} {\bibfnamefont
  {S.}~\bibnamefont {ichi Uchida}}, \bibinfo {author} {\bibfnamefont
  {A.}~\bibnamefont {Allais}}, \bibinfo {author} {\bibfnamefont {M.~J.}\
  \bibnamefont {Lawler}}, \bibinfo {author} {\bibfnamefont {E.-A.}\
  \bibnamefont {Kim}}, \bibinfo {author} {\bibfnamefont {S.}~\bibnamefont
  {Sachdev}},\ and\ \bibinfo {author} {\bibfnamefont {J.~C.~S.}\ \bibnamefont
  {Davis}},\ }\bibfield  {title} {\bibinfo {title} {Direct phase-sensitive
  identification of a d-form factor density wave in underdoped cuprates},\
  }\href {https://doi.org/10.1073/pnas.1406297111} {\bibfield  {journal}
  {\bibinfo  {journal} {Proc. Nat. Acad. Sci.}\ }\textbf {\bibinfo {volume}
  {111}},\ \bibinfo {pages} {E3026} (\bibinfo {year} {2014})}\BibitemShut
  {NoStop}%
\bibitem [{\citenamefont {Gerber}\ \emph {et~al.}(2015)\citenamefont {Gerber},
  \citenamefont {Jang}, \citenamefont {Nojiri}, \citenamefont {Matsuzawa},
  \citenamefont {Yasumura}, \citenamefont {Bonn}, \citenamefont {Liang},
  \citenamefont {Hardy}, \citenamefont {Islam}, \citenamefont {Mehta},
  \citenamefont {Song}, \citenamefont {Sikorski}, \citenamefont {Stefanescu},
  \citenamefont {Feng}, \citenamefont {Kivelson}, \citenamefont {Devereaux},
  \citenamefont {Shen}, \citenamefont {Kao}, \citenamefont {Lee}, \citenamefont
  {Zhu},\ and\ \citenamefont {Lee}}]{Gerber2015}%
  \BibitemOpen
  \bibfield  {author} {\bibinfo {author} {\bibfnamefont {S.}~\bibnamefont
  {Gerber}}, \bibinfo {author} {\bibfnamefont {H.}~\bibnamefont {Jang}},
  \bibinfo {author} {\bibfnamefont {H.}~\bibnamefont {Nojiri}}, \bibinfo
  {author} {\bibfnamefont {S.}~\bibnamefont {Matsuzawa}}, \bibinfo {author}
  {\bibfnamefont {H.}~\bibnamefont {Yasumura}}, \bibinfo {author}
  {\bibfnamefont {D.~A.}\ \bibnamefont {Bonn}}, \bibinfo {author}
  {\bibfnamefont {R.}~\bibnamefont {Liang}}, \bibinfo {author} {\bibfnamefont
  {W.~N.}\ \bibnamefont {Hardy}}, \bibinfo {author} {\bibfnamefont
  {Z.}~\bibnamefont {Islam}}, \bibinfo {author} {\bibfnamefont
  {A.}~\bibnamefont {Mehta}}, \bibinfo {author} {\bibfnamefont
  {S.}~\bibnamefont {Song}}, \bibinfo {author} {\bibfnamefont {M.}~\bibnamefont
  {Sikorski}}, \bibinfo {author} {\bibfnamefont {D.}~\bibnamefont
  {Stefanescu}}, \bibinfo {author} {\bibfnamefont {Y.}~\bibnamefont {Feng}},
  \bibinfo {author} {\bibfnamefont {S.~A.}\ \bibnamefont {Kivelson}}, \bibinfo
  {author} {\bibfnamefont {T.~P.}\ \bibnamefont {Devereaux}}, \bibinfo {author}
  {\bibfnamefont {Z.-X.}\ \bibnamefont {Shen}}, \bibinfo {author}
  {\bibfnamefont {C.-C.}\ \bibnamefont {Kao}}, \bibinfo {author} {\bibfnamefont
  {W.-S.}\ \bibnamefont {Lee}}, \bibinfo {author} {\bibfnamefont
  {D.}~\bibnamefont {Zhu}},\ and\ \bibinfo {author} {\bibfnamefont {J.-S.}\
  \bibnamefont {Lee}},\ }\bibfield  {title} {\bibinfo {title}
  {Three-dimensional charge density wave order in {YBa$_2$Cu$_3$O$_{6.67}$} at
  high magnetic fields},\ }\href {https://doi.org/10.1126/science.aac6257}
  {\bibfield  {journal} {\bibinfo  {journal} {Science}\ }\textbf {\bibinfo
  {volume} {350}},\ \bibinfo {pages} {949} (\bibinfo {year}
  {2015})}\BibitemShut {NoStop}%
\bibitem [{\citenamefont {Cheng}\ \emph {et~al.}(2022)\citenamefont {Cheng},
  \citenamefont {Zong}, \citenamefont {Li}, \citenamefont {Xia}, \citenamefont
  {Duan}, \citenamefont {Zhao}, \citenamefont {Li}, \citenamefont {Qi},
  \citenamefont {Wu}, \citenamefont {Zhao}, \citenamefont {Zhu}, \citenamefont
  {Zou}, \citenamefont {Jiang}, \citenamefont {Guo}, \citenamefont {Yang},
  \citenamefont {Qian}, \citenamefont {Zhang}, \citenamefont {Kogar},
  \citenamefont {Zuerch}, \citenamefont {Xiang},\ and\ \citenamefont
  {Zhang}}]{Cheng2022}%
  \BibitemOpen
  \bibfield  {author} {\bibinfo {author} {\bibfnamefont {Y.}~\bibnamefont
  {Cheng}}, \bibinfo {author} {\bibfnamefont {A.}~\bibnamefont {Zong}},
  \bibinfo {author} {\bibfnamefont {J.}~\bibnamefont {Li}}, \bibinfo {author}
  {\bibfnamefont {W.}~\bibnamefont {Xia}}, \bibinfo {author} {\bibfnamefont
  {S.}~\bibnamefont {Duan}}, \bibinfo {author} {\bibfnamefont {W.}~\bibnamefont
  {Zhao}}, \bibinfo {author} {\bibfnamefont {Y.}~\bibnamefont {Li}}, \bibinfo
  {author} {\bibfnamefont {F.}~\bibnamefont {Qi}}, \bibinfo {author}
  {\bibfnamefont {J.}~\bibnamefont {Wu}}, \bibinfo {author} {\bibfnamefont
  {L.}~\bibnamefont {Zhao}}, \bibinfo {author} {\bibfnamefont {P.}~\bibnamefont
  {Zhu}}, \bibinfo {author} {\bibfnamefont {X.}~\bibnamefont {Zou}}, \bibinfo
  {author} {\bibfnamefont {T.}~\bibnamefont {Jiang}}, \bibinfo {author}
  {\bibfnamefont {Y.}~\bibnamefont {Guo}}, \bibinfo {author} {\bibfnamefont
  {L.}~\bibnamefont {Yang}}, \bibinfo {author} {\bibfnamefont {D.}~\bibnamefont
  {Qian}}, \bibinfo {author} {\bibfnamefont {W.}~\bibnamefont {Zhang}},
  \bibinfo {author} {\bibfnamefont {A.}~\bibnamefont {Kogar}}, \bibinfo
  {author} {\bibfnamefont {M.~W.}\ \bibnamefont {Zuerch}}, \bibinfo {author}
  {\bibfnamefont {D.}~\bibnamefont {Xiang}},\ and\ \bibinfo {author}
  {\bibfnamefont {J.}~\bibnamefont {Zhang}},\ }\bibfield  {title} {\bibinfo
  {title} {Light-induced dimension crossover dictated by excitonic
  correlations},\ }\href {https://doi.org/10.1038/s41467-022-28309-5}
  {\bibfield  {journal} {\bibinfo  {journal} {Nat. Commun.}\ }\textbf {\bibinfo
  {volume} {13}},\ \bibinfo {pages} {963} (\bibinfo {year} {2022})}\BibitemShut
  {NoStop}%
\bibitem [{\citenamefont {Gao}\ \emph {et~al.}(2024)\citenamefont {Gao},
  \citenamefont {Chan}, \citenamefont {Jiao}, \citenamefont {Chen},
  \citenamefont {Yin}, \citenamefont {Tangprapha}, \citenamefont {Yang},
  \citenamefont {Li}, \citenamefont {Liu}, \citenamefont {Shen}, \citenamefont
  {Jiang},\ and\ \citenamefont {Chen}}]{Gao2024}%
  \BibitemOpen
  \bibfield  {author} {\bibinfo {author} {\bibfnamefont {Q.}~\bibnamefont
  {Gao}}, \bibinfo {author} {\bibfnamefont {Y.-h.}\ \bibnamefont {Chan}},
  \bibinfo {author} {\bibfnamefont {P.}~\bibnamefont {Jiao}}, \bibinfo {author}
  {\bibfnamefont {H.}~\bibnamefont {Chen}}, \bibinfo {author} {\bibfnamefont
  {S.}~\bibnamefont {Yin}}, \bibinfo {author} {\bibfnamefont {K.}~\bibnamefont
  {Tangprapha}}, \bibinfo {author} {\bibfnamefont {Y.}~\bibnamefont {Yang}},
  \bibinfo {author} {\bibfnamefont {X.}~\bibnamefont {Li}}, \bibinfo {author}
  {\bibfnamefont {Z.}~\bibnamefont {Liu}}, \bibinfo {author} {\bibfnamefont
  {D.}~\bibnamefont {Shen}}, \bibinfo {author} {\bibfnamefont {S.}~\bibnamefont
  {Jiang}},\ and\ \bibinfo {author} {\bibfnamefont {P.}~\bibnamefont {Chen}},\
  }\bibfield  {title} {\bibinfo {title} {Observation of possible excitonic
  charge density waves and metal-insulator transitions in atomically thin
  semimetals},\ }\href {https://doi.org/10.1038/s41567-023-02349-0} {\bibfield
  {journal} {\bibinfo  {journal} {Nat. Phys.}\ }\textbf {\bibinfo {volume}
  {20}},\ \bibinfo {pages} {597} (\bibinfo {year} {2024})}\BibitemShut
  {NoStop}%
\bibitem [{\citenamefont {Song}\ \emph {et~al.}(2023)\citenamefont {Song},
  \citenamefont {Jia}, \citenamefont {Xiong}, \citenamefont {Wang},
  \citenamefont {Jiang}, \citenamefont {Huang}, \citenamefont {Hwang},
  \citenamefont {Li}, \citenamefont {Hwang}, \citenamefont {Liu}, \citenamefont
  {Shen}, \citenamefont {Sobota}, \citenamefont {Kirchmann}, \citenamefont
  {Xue}, \citenamefont {Devereaux}, \citenamefont {Mo}, \citenamefont {Shen},\
  and\ \citenamefont {Tang}}]{Song2023}%
  \BibitemOpen
  \bibfield  {author} {\bibinfo {author} {\bibfnamefont {Y.}~\bibnamefont
  {Song}}, \bibinfo {author} {\bibfnamefont {C.}~\bibnamefont {Jia}}, \bibinfo
  {author} {\bibfnamefont {H.}~\bibnamefont {Xiong}}, \bibinfo {author}
  {\bibfnamefont {B.}~\bibnamefont {Wang}}, \bibinfo {author} {\bibfnamefont
  {Z.}~\bibnamefont {Jiang}}, \bibinfo {author} {\bibfnamefont
  {K.}~\bibnamefont {Huang}}, \bibinfo {author} {\bibfnamefont
  {J.}~\bibnamefont {Hwang}}, \bibinfo {author} {\bibfnamefont
  {Z.}~\bibnamefont {Li}}, \bibinfo {author} {\bibfnamefont {C.}~\bibnamefont
  {Hwang}}, \bibinfo {author} {\bibfnamefont {Z.}~\bibnamefont {Liu}}, \bibinfo
  {author} {\bibfnamefont {D.}~\bibnamefont {Shen}}, \bibinfo {author}
  {\bibfnamefont {J.~A.}\ \bibnamefont {Sobota}}, \bibinfo {author}
  {\bibfnamefont {P.}~\bibnamefont {Kirchmann}}, \bibinfo {author}
  {\bibfnamefont {J.}~\bibnamefont {Xue}}, \bibinfo {author} {\bibfnamefont
  {T.~P.}\ \bibnamefont {Devereaux}}, \bibinfo {author} {\bibfnamefont {S.-K.}\
  \bibnamefont {Mo}}, \bibinfo {author} {\bibfnamefont {Z.-X.}\ \bibnamefont
  {Shen}},\ and\ \bibinfo {author} {\bibfnamefont {S.}~\bibnamefont {Tang}},\
  }\bibfield  {title} {\bibinfo {title} {Signatures of the exciton gas phase
  and its condensation in monolayer {1T-ZrTe$_2$}},\ }\href
  {https://doi.org/10.1038/s41467-023-36857-7} {\bibfield  {journal} {\bibinfo
  {journal} {Nat. Commun.}\ }\textbf {\bibinfo {volume} {14}},\ \bibinfo
  {pages} {1116} (\bibinfo {year} {2023})}\BibitemShut {NoStop}%
\bibitem [{\citenamefont {Bao}\ \emph {et~al.}(2022)\citenamefont {Bao},
  \citenamefont {Tang}, \citenamefont {Sun},\ and\ \citenamefont
  {Zhou}}]{Bao2022}%
  \BibitemOpen
  \bibfield  {author} {\bibinfo {author} {\bibfnamefont {C.}~\bibnamefont
  {Bao}}, \bibinfo {author} {\bibfnamefont {P.}~\bibnamefont {Tang}}, \bibinfo
  {author} {\bibfnamefont {D.}~\bibnamefont {Sun}},\ and\ \bibinfo {author}
  {\bibfnamefont {S.}~\bibnamefont {Zhou}},\ }\bibfield  {title} {\bibinfo
  {title} {Light-induced emergent phenomena in 2{D} materials and topological
  materials},\ }\href {https://doi.org/10.1038/s42254-021-00388-1} {\bibfield
  {journal} {\bibinfo  {journal} {Nat. Rev. Phys.}\ }\textbf {\bibinfo {volume}
  {4}},\ \bibinfo {pages} {33} (\bibinfo {year} {2022})}\BibitemShut {NoStop}%
\bibitem [{\citenamefont {de~la Torre}\ \emph {et~al.}(2021)\citenamefont
  {de~la Torre}, \citenamefont {Kennes}, \citenamefont {Claassen},
  \citenamefont {Gerber}, \citenamefont {McIver},\ and\ \citenamefont
  {Sentef}}]{Torre2021}%
  \BibitemOpen
  \bibfield  {author} {\bibinfo {author} {\bibfnamefont {A.}~\bibnamefont
  {de~la Torre}}, \bibinfo {author} {\bibfnamefont {D.~M.}\ \bibnamefont
  {Kennes}}, \bibinfo {author} {\bibfnamefont {M.}~\bibnamefont {Claassen}},
  \bibinfo {author} {\bibfnamefont {S.}~\bibnamefont {Gerber}}, \bibinfo
  {author} {\bibfnamefont {J.~W.}\ \bibnamefont {McIver}},\ and\ \bibinfo
  {author} {\bibfnamefont {M.~A.}\ \bibnamefont {Sentef}},\ }\bibfield  {title}
  {\bibinfo {title} {Colloquium: Nonthermal pathways to ultrafast control in
  quantum materials},\ }\href {https://doi.org/10.1103/RevModPhys.93.041002}
  {\bibfield  {journal} {\bibinfo  {journal} {Rev. Mod. Phys.}\ }\textbf
  {\bibinfo {volume} {93}},\ \bibinfo {pages} {041002} (\bibinfo {year}
  {2021})}\BibitemShut {NoStop}%
\bibitem [{\citenamefont {Xu}\ \emph {et~al.}(2022)\citenamefont {Xu},
  \citenamefont {Chen},\ and\ \citenamefont {Meng}}]{Xu2022}%
  \BibitemOpen
  \bibfield  {author} {\bibinfo {author} {\bibfnamefont {J.}~\bibnamefont
  {Xu}}, \bibinfo {author} {\bibfnamefont {D.}~\bibnamefont {Chen}},\ and\
  \bibinfo {author} {\bibfnamefont {S.}~\bibnamefont {Meng}},\ }\bibfield
  {title} {\bibinfo {title} {Decoupled ultrafast electronic and structural
  phase transitions in photoexcited monoclinic {VO$_2$}},\ }\href
  {https://doi.org/10.1126/sciadv.add2392} {\bibfield  {journal} {\bibinfo
  {journal} {Sci. Adv.}\ }\textbf {\bibinfo {volume} {8}},\ \bibinfo {pages}
  {eadd2392} (\bibinfo {year} {2022})}\BibitemShut {NoStop}%
\bibitem [{\citenamefont {Perfetti}\ \emph {et~al.}(2007)\citenamefont
  {Perfetti}, \citenamefont {Loukakos}, \citenamefont {Lisowski}, \citenamefont
  {Bovensiepen}, \citenamefont {Eisaki},\ and\ \citenamefont
  {Wolf}}]{Perfetti2007}%
  \BibitemOpen
  \bibfield  {author} {\bibinfo {author} {\bibfnamefont {L.}~\bibnamefont
  {Perfetti}}, \bibinfo {author} {\bibfnamefont {P.~A.}\ \bibnamefont
  {Loukakos}}, \bibinfo {author} {\bibfnamefont {M.}~\bibnamefont {Lisowski}},
  \bibinfo {author} {\bibfnamefont {U.}~\bibnamefont {Bovensiepen}}, \bibinfo
  {author} {\bibfnamefont {H.}~\bibnamefont {Eisaki}},\ and\ \bibinfo {author}
  {\bibfnamefont {M.}~\bibnamefont {Wolf}},\ }\bibfield  {title} {\bibinfo
  {title} {Ultrafast electron relaxation in superconducting
  {${\mathrm{Bi}}_{2}{\mathrm{Sr}}_{2}{\mathrm{CaCu}}_{2}{\mathrm{O}}_{8+\ensuremath{\delta}}$}
  by time-resolved photoelectron spectroscopy},\ }\href
  {https://doi.org/10.1103/PhysRevLett.99.197001} {\bibfield  {journal}
  {\bibinfo  {journal} {Phys. Rev. Lett.}\ }\textbf {\bibinfo {volume} {99}},\
  \bibinfo {pages} {197001} (\bibinfo {year} {2007})}\BibitemShut {NoStop}%
\bibitem [{\citenamefont {Katsumi}\ \emph {et~al.}(2023)\citenamefont
  {Katsumi}, \citenamefont {Alekhin}, \citenamefont {Souliou}, \citenamefont
  {Merz}, \citenamefont {Haghighirad}, \citenamefont {Le~Tacon}, \citenamefont
  {Houver}, \citenamefont {Cazayous}, \citenamefont {Sacuto},\ and\
  \citenamefont {Gallais}}]{Katsumi2023}%
  \BibitemOpen
  \bibfield  {author} {\bibinfo {author} {\bibfnamefont {K.}~\bibnamefont
  {Katsumi}}, \bibinfo {author} {\bibfnamefont {A.}~\bibnamefont {Alekhin}},
  \bibinfo {author} {\bibfnamefont {S.-M.}\ \bibnamefont {Souliou}}, \bibinfo
  {author} {\bibfnamefont {M.}~\bibnamefont {Merz}}, \bibinfo {author}
  {\bibfnamefont {A.-A.}\ \bibnamefont {Haghighirad}}, \bibinfo {author}
  {\bibfnamefont {M.}~\bibnamefont {Le~Tacon}}, \bibinfo {author}
  {\bibfnamefont {S.}~\bibnamefont {Houver}}, \bibinfo {author} {\bibfnamefont
  {M.}~\bibnamefont {Cazayous}}, \bibinfo {author} {\bibfnamefont
  {A.}~\bibnamefont {Sacuto}},\ and\ \bibinfo {author} {\bibfnamefont
  {Y.}~\bibnamefont {Gallais}},\ }\bibfield  {title} {\bibinfo {title}
  {Disentangling lattice and electronic instabilities in the excitonic
  insulator candidate {${\mathrm{Ta}}_{2}{\mathrm{NiSe}}_{5}$} by
  nonequilibrium spectroscopy},\ }\href
  {https://doi.org/10.1103/PhysRevLett.130.106904} {\bibfield  {journal}
  {\bibinfo  {journal} {Phys. Rev. Lett.}\ }\textbf {\bibinfo {volume} {130}},\
  \bibinfo {pages} {106904} (\bibinfo {year} {2023})}\BibitemShut {NoStop}%
\bibitem [{\citenamefont {Tang}\ \emph {et~al.}(2020)\citenamefont {Tang},
  \citenamefont {Wang}, \citenamefont {Duan}, \citenamefont {Yang},
  \citenamefont {Huang}, \citenamefont {Guo}, \citenamefont {Qian},\ and\
  \citenamefont {Zhang}}]{Tang2020}%
  \BibitemOpen
  \bibfield  {author} {\bibinfo {author} {\bibfnamefont {T.}~\bibnamefont
  {Tang}}, \bibinfo {author} {\bibfnamefont {H.}~\bibnamefont {Wang}}, \bibinfo
  {author} {\bibfnamefont {S.}~\bibnamefont {Duan}}, \bibinfo {author}
  {\bibfnamefont {Y.}~\bibnamefont {Yang}}, \bibinfo {author} {\bibfnamefont
  {C.}~\bibnamefont {Huang}}, \bibinfo {author} {\bibfnamefont
  {Y.}~\bibnamefont {Guo}}, \bibinfo {author} {\bibfnamefont {D.}~\bibnamefont
  {Qian}},\ and\ \bibinfo {author} {\bibfnamefont {W.}~\bibnamefont {Zhang}},\
  }\bibfield  {title} {\bibinfo {title} {Non-coulomb strong electron-hole
  binding in {${\mathrm{Ta}}_{2}{\mathrm{NiSe}}_{5}$} revealed by time- and
  angle-resolved photoemission spectroscopy},\ }\href
  {https://doi.org/10.1103/PhysRevB.101.235148} {\bibfield  {journal} {\bibinfo
   {journal} {Phys. Rev. B}\ }\textbf {\bibinfo {volume} {101}},\ \bibinfo
  {pages} {235148} (\bibinfo {year} {2020})}\BibitemShut {NoStop}%
\bibitem [{\citenamefont {Rohwer}\ \emph {et~al.}(2011)\citenamefont {Rohwer},
  \citenamefont {Hellmann}, \citenamefont {Wiesenmayer}, \citenamefont {Sohrt},
  \citenamefont {Stange}, \citenamefont {Slomski}, \citenamefont {Carr},
  \citenamefont {Liu}, \citenamefont {Avila}, \citenamefont {Kall$\ddot{a}$ne},
  \citenamefont {Mathias}, \citenamefont {Kipp}, \citenamefont {Rossnagel},\
  and\ \citenamefont {Bauer}}]{Rohwer2011}%
  \BibitemOpen
  \bibfield  {author} {\bibinfo {author} {\bibfnamefont {T.}~\bibnamefont
  {Rohwer}}, \bibinfo {author} {\bibfnamefont {S.}~\bibnamefont {Hellmann}},
  \bibinfo {author} {\bibfnamefont {M.}~\bibnamefont {Wiesenmayer}}, \bibinfo
  {author} {\bibfnamefont {C.}~\bibnamefont {Sohrt}}, \bibinfo {author}
  {\bibfnamefont {A.}~\bibnamefont {Stange}}, \bibinfo {author} {\bibfnamefont
  {B.}~\bibnamefont {Slomski}}, \bibinfo {author} {\bibfnamefont
  {A.}~\bibnamefont {Carr}}, \bibinfo {author} {\bibfnamefont {Y.}~\bibnamefont
  {Liu}}, \bibinfo {author} {\bibfnamefont {L.~M.}\ \bibnamefont {Avila}},
  \bibinfo {author} {\bibfnamefont {M.}~\bibnamefont {Kall$\ddot{a}$ne}},
  \bibinfo {author} {\bibfnamefont {S.}~\bibnamefont {Mathias}}, \bibinfo
  {author} {\bibfnamefont {L.}~\bibnamefont {Kipp}}, \bibinfo {author}
  {\bibfnamefont {K.}~\bibnamefont {Rossnagel}},\ and\ \bibinfo {author}
  {\bibfnamefont {M.}~\bibnamefont {Bauer}},\ }\bibfield  {title} {\bibinfo
  {title} {Collapse of long-range charge order tracked by time-resolved
  photoemission at high momenta},\ }\href {https://doi.org/10.1038/nature09829}
  {\bibfield  {journal} {\bibinfo  {journal} {Nature}\ }\textbf {\bibinfo
  {volume} {471}},\ \bibinfo {pages} {490} (\bibinfo {year}
  {2011})}\BibitemShut {NoStop}%
\bibitem [{\citenamefont {Maklar}\ \emph {et~al.}(2021)\citenamefont {Maklar},
  \citenamefont {Windsor}, \citenamefont {Nicholson}, \citenamefont {Puppin},
  \citenamefont {Walmsley}, \citenamefont {Esposito}, \citenamefont {Porer},
  \citenamefont {Rittmann}, \citenamefont {Leuenberger}, \citenamefont {Kubli},
  \citenamefont {Savoini}, \citenamefont {Abreu}, \citenamefont {Johnson},
  \citenamefont {Beaud}, \citenamefont {Ingold}, \citenamefont {Staub},
  \citenamefont {Fisher}, \citenamefont {Ernstorfer}, \citenamefont {Wolf},\
  and\ \citenamefont {Rettig}}]{Maklar2021}%
  \BibitemOpen
  \bibfield  {author} {\bibinfo {author} {\bibfnamefont {J.}~\bibnamefont
  {Maklar}}, \bibinfo {author} {\bibfnamefont {Y.~W.}\ \bibnamefont {Windsor}},
  \bibinfo {author} {\bibfnamefont {C.~W.}\ \bibnamefont {Nicholson}}, \bibinfo
  {author} {\bibfnamefont {M.}~\bibnamefont {Puppin}}, \bibinfo {author}
  {\bibfnamefont {P.}~\bibnamefont {Walmsley}}, \bibinfo {author}
  {\bibfnamefont {V.}~\bibnamefont {Esposito}}, \bibinfo {author}
  {\bibfnamefont {M.}~\bibnamefont {Porer}}, \bibinfo {author} {\bibfnamefont
  {J.}~\bibnamefont {Rittmann}}, \bibinfo {author} {\bibfnamefont
  {D.}~\bibnamefont {Leuenberger}}, \bibinfo {author} {\bibfnamefont
  {M.}~\bibnamefont {Kubli}}, \bibinfo {author} {\bibfnamefont
  {M.}~\bibnamefont {Savoini}}, \bibinfo {author} {\bibfnamefont
  {E.}~\bibnamefont {Abreu}}, \bibinfo {author} {\bibfnamefont {S.~L.}\
  \bibnamefont {Johnson}}, \bibinfo {author} {\bibfnamefont {P.}~\bibnamefont
  {Beaud}}, \bibinfo {author} {\bibfnamefont {G.}~\bibnamefont {Ingold}},
  \bibinfo {author} {\bibfnamefont {U.}~\bibnamefont {Staub}}, \bibinfo
  {author} {\bibfnamefont {I.~R.}\ \bibnamefont {Fisher}}, \bibinfo {author}
  {\bibfnamefont {R.}~\bibnamefont {Ernstorfer}}, \bibinfo {author}
  {\bibfnamefont {M.}~\bibnamefont {Wolf}},\ and\ \bibinfo {author}
  {\bibfnamefont {L.}~\bibnamefont {Rettig}},\ }\bibfield  {title} {\bibinfo
  {title} {Nonequilibrium charge-density-wave order beyond the thermal limit},\
  }\href {https://doi.org/10.1038/s41467-021-22778-w} {\bibfield  {journal}
  {\bibinfo  {journal} {Nat. Commun.}\ }\textbf {\bibinfo {volume} {12}},\
  \bibinfo {pages} {2499} (\bibinfo {year} {2021})}\BibitemShut {NoStop}%
\bibitem [{\citenamefont {Yang}\ \emph {et~al.}(2022)\citenamefont {Yang},
  \citenamefont {Wang}, \citenamefont {Duan}, \citenamefont {Wo}, \citenamefont
  {Huang}, \citenamefont {Wang}, \citenamefont {Gu}, \citenamefont {Xiang},
  \citenamefont {Qian}, \citenamefont {Zhao},\ and\ \citenamefont
  {Zhang}}]{Yang2022}%
  \BibitemOpen
  \bibfield  {author} {\bibinfo {author} {\bibfnamefont {Y.}~\bibnamefont
  {Yang}}, \bibinfo {author} {\bibfnamefont {Q.}~\bibnamefont {Wang}}, \bibinfo
  {author} {\bibfnamefont {S.}~\bibnamefont {Duan}}, \bibinfo {author}
  {\bibfnamefont {H.}~\bibnamefont {Wo}}, \bibinfo {author} {\bibfnamefont
  {C.}~\bibnamefont {Huang}}, \bibinfo {author} {\bibfnamefont
  {S.}~\bibnamefont {Wang}}, \bibinfo {author} {\bibfnamefont {L.}~\bibnamefont
  {Gu}}, \bibinfo {author} {\bibfnamefont {D.}~\bibnamefont {Xiang}}, \bibinfo
  {author} {\bibfnamefont {D.}~\bibnamefont {Qian}}, \bibinfo {author}
  {\bibfnamefont {J.}~\bibnamefont {Zhao}},\ and\ \bibinfo {author}
  {\bibfnamefont {W.}~\bibnamefont {Zhang}},\ }\bibfield  {title} {\bibinfo
  {title} {Anomalous contribution to the nematic electronic states from the
  structural transition in fese revealed by time- and angle-resolved
  photoemission spectroscopy},\ }\href
  {https://doi.org/10.1103/PhysRevLett.128.246401} {\bibfield  {journal}
  {\bibinfo  {journal} {Phys. Rev. Lett.}\ }\textbf {\bibinfo {volume} {128}},\
  \bibinfo {pages} {246401} (\bibinfo {year} {2022})}\BibitemShut {NoStop}%
\bibitem [{\citenamefont {Kogar}\ \emph {et~al.}(2020)\citenamefont {Kogar},
  \citenamefont {Zong}, \citenamefont {Dolgirev}, \citenamefont {Shen},
  \citenamefont {Straquadine}, \citenamefont {Bie}, \citenamefont {Wang},
  \citenamefont {Rohwer}, \citenamefont {Tung}, \citenamefont {Yang},
  \citenamefont {Li}, \citenamefont {Yang}, \citenamefont {Weathersby},
  \citenamefont {Park}, \citenamefont {Kozina}, \citenamefont {Sie},
  \citenamefont {Wen}, \citenamefont {Jarillo-Herrero}, \citenamefont {Fisher},
  \citenamefont {Wang},\ and\ \citenamefont {Gedik}}]{Kogar2020}%
  \BibitemOpen
  \bibfield  {author} {\bibinfo {author} {\bibfnamefont {A.}~\bibnamefont
  {Kogar}}, \bibinfo {author} {\bibfnamefont {A.}~\bibnamefont {Zong}},
  \bibinfo {author} {\bibfnamefont {P.~E.}\ \bibnamefont {Dolgirev}}, \bibinfo
  {author} {\bibfnamefont {X.}~\bibnamefont {Shen}}, \bibinfo {author}
  {\bibfnamefont {J.}~\bibnamefont {Straquadine}}, \bibinfo {author}
  {\bibfnamefont {Y.-Q.}\ \bibnamefont {Bie}}, \bibinfo {author} {\bibfnamefont
  {X.}~\bibnamefont {Wang}}, \bibinfo {author} {\bibfnamefont {T.}~\bibnamefont
  {Rohwer}}, \bibinfo {author} {\bibfnamefont {I.-C.}\ \bibnamefont {Tung}},
  \bibinfo {author} {\bibfnamefont {Y.}~\bibnamefont {Yang}}, \bibinfo {author}
  {\bibfnamefont {R.}~\bibnamefont {Li}}, \bibinfo {author} {\bibfnamefont
  {J.}~\bibnamefont {Yang}}, \bibinfo {author} {\bibfnamefont {S.}~\bibnamefont
  {Weathersby}}, \bibinfo {author} {\bibfnamefont {S.}~\bibnamefont {Park}},
  \bibinfo {author} {\bibfnamefont {M.~E.}\ \bibnamefont {Kozina}}, \bibinfo
  {author} {\bibfnamefont {E.~J.}\ \bibnamefont {Sie}}, \bibinfo {author}
  {\bibfnamefont {H.}~\bibnamefont {Wen}}, \bibinfo {author} {\bibfnamefont
  {P.}~\bibnamefont {Jarillo-Herrero}}, \bibinfo {author} {\bibfnamefont
  {I.~R.}\ \bibnamefont {Fisher}}, \bibinfo {author} {\bibfnamefont
  {X.}~\bibnamefont {Wang}},\ and\ \bibinfo {author} {\bibfnamefont
  {N.}~\bibnamefont {Gedik}},\ }\bibfield  {title} {\bibinfo {title}
  {Light-induced charge density wave in {LaTe$_3$}},\ }\href
  {https://doi.org/10.1038/s41567-019-0705-3} {\bibfield  {journal} {\bibinfo
  {journal} {Nat. Phys.}\ }\textbf {\bibinfo {volume} {16}},\ \bibinfo {pages}
  {159} (\bibinfo {year} {2020})}\BibitemShut {NoStop}%
\bibitem [{\citenamefont {Gerasimenko}\ \emph {et~al.}(2019)\citenamefont
  {Gerasimenko}, \citenamefont {Karpov}, \citenamefont {Vaskivskyi},
  \citenamefont {Brazovskii},\ and\ \citenamefont
  {Mihailovic}}]{Gerasimenko2019}%
  \BibitemOpen
  \bibfield  {author} {\bibinfo {author} {\bibfnamefont {Y.~A.}\ \bibnamefont
  {Gerasimenko}}, \bibinfo {author} {\bibfnamefont {P.}~\bibnamefont {Karpov}},
  \bibinfo {author} {\bibfnamefont {I.}~\bibnamefont {Vaskivskyi}}, \bibinfo
  {author} {\bibfnamefont {S.}~\bibnamefont {Brazovskii}},\ and\ \bibinfo
  {author} {\bibfnamefont {D.}~\bibnamefont {Mihailovic}},\ }\bibfield  {title}
  {\bibinfo {title} {Intertwined chiral charge orders and topological
  stabilization of the light-induced state of a prototypical transition metal
  dichalcogenide},\ }\href {https://doi.org/10.1038/s41535-019-0172-1}
  {\bibfield  {journal} {\bibinfo  {journal} {npj Quantum Materials}\ }\textbf
  {\bibinfo {volume} {4}},\ \bibinfo {pages} {32} (\bibinfo {year}
  {2019})}\BibitemShut {NoStop}%
\bibitem [{\citenamefont {Duan}\ \emph {et~al.}(2021)\citenamefont {Duan},
  \citenamefont {Cheng}, \citenamefont {Xia}, \citenamefont {Yang},
  \citenamefont {Xu}, \citenamefont {Qi}, \citenamefont {Huang}, \citenamefont
  {Tang}, \citenamefont {Guo}, \citenamefont {Luo}, \citenamefont {Qian},
  \citenamefont {Xiang}, \citenamefont {Zhang},\ and\ \citenamefont
  {Zhang}}]{Duan2021}%
  \BibitemOpen
  \bibfield  {author} {\bibinfo {author} {\bibfnamefont {S.}~\bibnamefont
  {Duan}}, \bibinfo {author} {\bibfnamefont {Y.}~\bibnamefont {Cheng}},
  \bibinfo {author} {\bibfnamefont {W.}~\bibnamefont {Xia}}, \bibinfo {author}
  {\bibfnamefont {Y.}~\bibnamefont {Yang}}, \bibinfo {author} {\bibfnamefont
  {C.}~\bibnamefont {Xu}}, \bibinfo {author} {\bibfnamefont {F.}~\bibnamefont
  {Qi}}, \bibinfo {author} {\bibfnamefont {C.}~\bibnamefont {Huang}}, \bibinfo
  {author} {\bibfnamefont {T.}~\bibnamefont {Tang}}, \bibinfo {author}
  {\bibfnamefont {Y.}~\bibnamefont {Guo}}, \bibinfo {author} {\bibfnamefont
  {W.}~\bibnamefont {Luo}}, \bibinfo {author} {\bibfnamefont {D.}~\bibnamefont
  {Qian}}, \bibinfo {author} {\bibfnamefont {D.}~\bibnamefont {Xiang}},
  \bibinfo {author} {\bibfnamefont {J.}~\bibnamefont {Zhang}},\ and\ \bibinfo
  {author} {\bibfnamefont {W.}~\bibnamefont {Zhang}},\ }\bibfield  {title}
  {\bibinfo {title} {Optical manipulation of electronic dimensionality in a
  quantum material},\ }\href {https://doi.org/10.1038/s41586-021-03643-8}
  {\bibfield  {journal} {\bibinfo  {journal} {Nature}\ }\textbf {\bibinfo
  {volume} {595}},\ \bibinfo {pages} {239} (\bibinfo {year}
  {2021})}\BibitemShut {NoStop}%
\bibitem [{\citenamefont {Yusupov}\ \emph {et~al.}(2010)\citenamefont
  {Yusupov}, \citenamefont {Mertelj}, \citenamefont {Kabanov}, \citenamefont
  {Brazovskii}, \citenamefont {Kusar}, \citenamefont {Chu}, \citenamefont
  {Fisher},\ and\ \citenamefont {Mihailovic}}]{Yusupov2010}%
  \BibitemOpen
  \bibfield  {author} {\bibinfo {author} {\bibfnamefont {R.}~\bibnamefont
  {Yusupov}}, \bibinfo {author} {\bibfnamefont {T.}~\bibnamefont {Mertelj}},
  \bibinfo {author} {\bibfnamefont {V.~V.}\ \bibnamefont {Kabanov}}, \bibinfo
  {author} {\bibfnamefont {S.}~\bibnamefont {Brazovskii}}, \bibinfo {author}
  {\bibfnamefont {P.}~\bibnamefont {Kusar}}, \bibinfo {author} {\bibfnamefont
  {J.-H.}\ \bibnamefont {Chu}}, \bibinfo {author} {\bibfnamefont {I.~R.}\
  \bibnamefont {Fisher}},\ and\ \bibinfo {author} {\bibfnamefont
  {D.}~\bibnamefont {Mihailovic}},\ }\bibfield  {title} {\bibinfo {title}
  {Coherent dynamics of macroscopic electronic order through a symmetry
  breaking transition},\ }\href {https://doi.org/10.1038/nphys1738} {\bibfield
  {journal} {\bibinfo  {journal} {Nat. Phys.}\ }\textbf {\bibinfo {volume}
  {6}},\ \bibinfo {pages} {681} (\bibinfo {year} {2010})}\BibitemShut {NoStop}%
\bibitem [{\citenamefont {Trigo}\ \emph {et~al.}(2021)\citenamefont {Trigo},
  \citenamefont {Giraldo-Gallo}, \citenamefont {Clark}, \citenamefont {Kozina},
  \citenamefont {Henighan}, \citenamefont {Jiang}, \citenamefont {Chollet},
  \citenamefont {Fisher}, \citenamefont {Glownia}, \citenamefont {Katayama},
  \citenamefont {Kirchmann}, \citenamefont {Leuenberger}, \citenamefont {Liu},
  \citenamefont {Reis}, \citenamefont {Shen},\ and\ \citenamefont
  {Zhu}}]{Trigo2021}%
  \BibitemOpen
  \bibfield  {author} {\bibinfo {author} {\bibfnamefont {M.}~\bibnamefont
  {Trigo}}, \bibinfo {author} {\bibfnamefont {P.}~\bibnamefont
  {Giraldo-Gallo}}, \bibinfo {author} {\bibfnamefont {J.~N.}\ \bibnamefont
  {Clark}}, \bibinfo {author} {\bibfnamefont {M.~E.}\ \bibnamefont {Kozina}},
  \bibinfo {author} {\bibfnamefont {T.}~\bibnamefont {Henighan}}, \bibinfo
  {author} {\bibfnamefont {M.~P.}\ \bibnamefont {Jiang}}, \bibinfo {author}
  {\bibfnamefont {M.}~\bibnamefont {Chollet}}, \bibinfo {author} {\bibfnamefont
  {I.~R.}\ \bibnamefont {Fisher}}, \bibinfo {author} {\bibfnamefont {J.~M.}\
  \bibnamefont {Glownia}}, \bibinfo {author} {\bibfnamefont {T.}~\bibnamefont
  {Katayama}}, \bibinfo {author} {\bibfnamefont {P.~S.}\ \bibnamefont
  {Kirchmann}}, \bibinfo {author} {\bibfnamefont {D.}~\bibnamefont
  {Leuenberger}}, \bibinfo {author} {\bibfnamefont {H.}~\bibnamefont {Liu}},
  \bibinfo {author} {\bibfnamefont {D.~A.}\ \bibnamefont {Reis}}, \bibinfo
  {author} {\bibfnamefont {Z.~X.}\ \bibnamefont {Shen}},\ and\ \bibinfo
  {author} {\bibfnamefont {D.}~\bibnamefont {Zhu}},\ }\bibfield  {title}
  {\bibinfo {title} {Ultrafast formation of domain walls of a charge density
  wave in {${\mathrm{SmTe}}_{3}$}},\ }\href
  {https://doi.org/10.1103/PhysRevB.103.054109} {\bibfield  {journal} {\bibinfo
   {journal} {Phys. Rev. B}\ }\textbf {\bibinfo {volume} {103}},\ \bibinfo
  {pages} {054109} (\bibinfo {year} {2021})}\BibitemShut {NoStop}%
\bibitem [{\citenamefont {Kong}\ \emph {et~al.}(2024)\citenamefont {Kong},
  \citenamefont {Shindou},\ and\ \citenamefont {Sun}}]{kong2024}%
  \BibitemOpen
  \bibfield  {author} {\bibinfo {author} {\bibfnamefont {L.}~\bibnamefont
  {Kong}}, \bibinfo {author} {\bibfnamefont {R.}~\bibnamefont {Shindou}},\ and\
  \bibinfo {author} {\bibfnamefont {Z.}~\bibnamefont {Sun}},\ }\bibfield
  {title} {\bibinfo {title} {Fate of transient order parameter domain walls in
  ultrafast experiments},\ }\href {https://doi.org/10.48550/arXiv.2407.14250}
  {\bibfield  {journal} {\bibinfo  {journal} {arXiv preprint arXiv:2407.14250}\
  } (\bibinfo {year} {2024})}\BibitemShut {NoStop}%
\bibitem [{\citenamefont {Duan}\ \emph {et~al.}(2023)\citenamefont {Duan},
  \citenamefont {Xia}, \citenamefont {Huang}, \citenamefont {Wang},
  \citenamefont {Gu}, \citenamefont {Liu}, \citenamefont {Xiang}, \citenamefont
  {Qian}, \citenamefont {Guo},\ and\ \citenamefont {Zhang}}]{Duan2023}%
  \BibitemOpen
  \bibfield  {author} {\bibinfo {author} {\bibfnamefont {S.}~\bibnamefont
  {Duan}}, \bibinfo {author} {\bibfnamefont {W.}~\bibnamefont {Xia}}, \bibinfo
  {author} {\bibfnamefont {C.}~\bibnamefont {Huang}}, \bibinfo {author}
  {\bibfnamefont {S.}~\bibnamefont {Wang}}, \bibinfo {author} {\bibfnamefont
  {L.}~\bibnamefont {Gu}}, \bibinfo {author} {\bibfnamefont {H.}~\bibnamefont
  {Liu}}, \bibinfo {author} {\bibfnamefont {D.}~\bibnamefont {Xiang}}, \bibinfo
  {author} {\bibfnamefont {D.}~\bibnamefont {Qian}}, \bibinfo {author}
  {\bibfnamefont {Y.}~\bibnamefont {Guo}},\ and\ \bibinfo {author}
  {\bibfnamefont {W.}~\bibnamefont {Zhang}},\ }\bibfield  {title} {\bibinfo
  {title} {Ultrafast switching from the charge density wave phase to a
  metastable metallic state in
  {$1T\text{\ensuremath{-}}{\mathrm{TiSe}}_{2}$}},\ }\href
  {https://doi.org/10.1103/PhysRevLett.130.226501} {\bibfield  {journal}
  {\bibinfo  {journal} {Phys. Rev. Lett.}\ }\textbf {\bibinfo {volume} {130}},\
  \bibinfo {pages} {226501} (\bibinfo {year} {2023})}\BibitemShut {NoStop}%
\bibitem [{\citenamefont {Brouet}\ \emph {et~al.}(2004)\citenamefont {Brouet},
  \citenamefont {Yang}, \citenamefont {Zhou}, \citenamefont {Hussain},
  \citenamefont {Ru}, \citenamefont {Shin}, \citenamefont {Fisher},\ and\
  \citenamefont {Shen}}]{Brouet2004}%
  \BibitemOpen
  \bibfield  {author} {\bibinfo {author} {\bibfnamefont {V.}~\bibnamefont
  {Brouet}}, \bibinfo {author} {\bibfnamefont {W.~L.}\ \bibnamefont {Yang}},
  \bibinfo {author} {\bibfnamefont {X.~J.}\ \bibnamefont {Zhou}}, \bibinfo
  {author} {\bibfnamefont {Z.}~\bibnamefont {Hussain}}, \bibinfo {author}
  {\bibfnamefont {N.}~\bibnamefont {Ru}}, \bibinfo {author} {\bibfnamefont
  {K.~Y.}\ \bibnamefont {Shin}}, \bibinfo {author} {\bibfnamefont {I.~R.}\
  \bibnamefont {Fisher}},\ and\ \bibinfo {author} {\bibfnamefont {Z.~X.}\
  \bibnamefont {Shen}},\ }\bibfield  {title} {\bibinfo {title} {Fermi surface
  reconstruction in the cdw state of {CeTe$_3$} observed by photoemission},\
  }\href {https://doi.org/10.1103/PhysRevLett.93.126405} {\bibfield  {journal}
  {\bibinfo  {journal} {Phys. Rev. Lett.}\ }\textbf {\bibinfo {volume} {93}},\
  \bibinfo {pages} {126405} (\bibinfo {year} {2004})}\BibitemShut {NoStop}%
\bibitem [{\citenamefont {Johannes}\ and\ \citenamefont
  {Mazin}(2008)}]{Johannes2008}%
  \BibitemOpen
  \bibfield  {author} {\bibinfo {author} {\bibfnamefont {M.~D.}\ \bibnamefont
  {Johannes}}\ and\ \bibinfo {author} {\bibfnamefont {I.~I.}\ \bibnamefont
  {Mazin}},\ }\bibfield  {title} {\bibinfo {title} {Fermi surface nesting and
  the origin of charge density waves in metals},\ }\href
  {https://doi.org/10.1103/PhysRevB.77.165135} {\bibfield  {journal} {\bibinfo
  {journal} {Phys. Rev. B}\ }\textbf {\bibinfo {volume} {77}},\ \bibinfo
  {pages} {165135} (\bibinfo {year} {2008})}\BibitemShut {NoStop}%
\bibitem [{\citenamefont {Maschek}\ \emph {et~al.}(2015)\citenamefont
  {Maschek}, \citenamefont {Rosenkranz}, \citenamefont {Heid}, \citenamefont
  {Said}, \citenamefont {Giraldo-Gallo}, \citenamefont {Fisher},\ and\
  \citenamefont {Weber}}]{Maschek2015}%
  \BibitemOpen
  \bibfield  {author} {\bibinfo {author} {\bibfnamefont {M.}~\bibnamefont
  {Maschek}}, \bibinfo {author} {\bibfnamefont {S.}~\bibnamefont {Rosenkranz}},
  \bibinfo {author} {\bibfnamefont {R.}~\bibnamefont {Heid}}, \bibinfo {author}
  {\bibfnamefont {A.~H.}\ \bibnamefont {Said}}, \bibinfo {author}
  {\bibfnamefont {P.}~\bibnamefont {Giraldo-Gallo}}, \bibinfo {author}
  {\bibfnamefont {I.~R.}\ \bibnamefont {Fisher}},\ and\ \bibinfo {author}
  {\bibfnamefont {F.}~\bibnamefont {Weber}},\ }\bibfield  {title} {\bibinfo
  {title} {Wave-vector-dependent electron-phonon coupling and the
  charge-density-wave transition in $\mathrm{TbT}{\mathrm{e}}_{3}$},\ }\href
  {https://doi.org/10.1103/PhysRevB.91.235146} {\bibfield  {journal} {\bibinfo
  {journal} {Phys. Rev. B}\ }\textbf {\bibinfo {volume} {91}},\ \bibinfo
  {pages} {235146} (\bibinfo {year} {2015})}\BibitemShut {NoStop}%
\bibitem [{\citenamefont {Eiter}\ \emph {et~al.}(2013)\citenamefont {Eiter},
  \citenamefont {Lavagnini}, \citenamefont {Hackl}, \citenamefont {Nowadnick},
  \citenamefont {Kemper}, \citenamefont {Devereaux}, \citenamefont {Chu},
  \citenamefont {Analytis}, \citenamefont {Fisher},\ and\ \citenamefont
  {Degiorgi}}]{Eiter2013}%
  \BibitemOpen
  \bibfield  {author} {\bibinfo {author} {\bibfnamefont {H.-M.}\ \bibnamefont
  {Eiter}}, \bibinfo {author} {\bibfnamefont {M.}~\bibnamefont {Lavagnini}},
  \bibinfo {author} {\bibfnamefont {R.}~\bibnamefont {Hackl}}, \bibinfo
  {author} {\bibfnamefont {E.~A.}\ \bibnamefont {Nowadnick}}, \bibinfo {author}
  {\bibfnamefont {A.~F.}\ \bibnamefont {Kemper}}, \bibinfo {author}
  {\bibfnamefont {T.~P.}\ \bibnamefont {Devereaux}}, \bibinfo {author}
  {\bibfnamefont {J.-H.}\ \bibnamefont {Chu}}, \bibinfo {author} {\bibfnamefont
  {J.~G.}\ \bibnamefont {Analytis}}, \bibinfo {author} {\bibfnamefont {I.~R.}\
  \bibnamefont {Fisher}},\ and\ \bibinfo {author} {\bibfnamefont
  {L.}~\bibnamefont {Degiorgi}},\ }\bibfield  {title} {\bibinfo {title}
  {Alternative route to charge density wave formation in multiband systems},\
  }\href {https://doi.org/10.1073/pnas.1214745110} {\bibfield  {journal}
  {\bibinfo  {journal} {Proc. Nat. Acad. Sci.}\ }\textbf {\bibinfo {volume}
  {110}},\ \bibinfo {pages} {64} (\bibinfo {year} {2013})}\BibitemShut
  {NoStop}%
\bibitem [{\citenamefont {Kim}\ \emph {et~al.}(2006)\citenamefont {Kim},
  \citenamefont {Malliakas}, \citenamefont {Tomi\ifmmode~\acute{c}\else
  \'{c}\fi{}}, \citenamefont {Tessmer}, \citenamefont {Kanatzidis},\ and\
  \citenamefont {Billinge}}]{Kim2006}%
  \BibitemOpen
  \bibfield  {author} {\bibinfo {author} {\bibfnamefont {H.~J.}\ \bibnamefont
  {Kim}}, \bibinfo {author} {\bibfnamefont {C.~D.}\ \bibnamefont {Malliakas}},
  \bibinfo {author} {\bibfnamefont {A.~T.}\ \bibnamefont
  {Tomi\ifmmode~\acute{c}\else \'{c}\fi{}}}, \bibinfo {author} {\bibfnamefont
  {S.~H.}\ \bibnamefont {Tessmer}}, \bibinfo {author} {\bibfnamefont {M.~G.}\
  \bibnamefont {Kanatzidis}},\ and\ \bibinfo {author} {\bibfnamefont
  {S.~J.~L.}\ \bibnamefont {Billinge}},\ }\bibfield  {title} {\bibinfo {title}
  {Local atomic structure and discommensurations in the charge density wave of
  {${\mathrm{CeTe}}_{3}$}},\ }\href
  {https://doi.org/10.1103/PhysRevLett.96.226401} {\bibfield  {journal}
  {\bibinfo  {journal} {Phys. Rev. Lett.}\ }\textbf {\bibinfo {volume} {96}},\
  \bibinfo {pages} {226401} (\bibinfo {year} {2006})}\BibitemShut {NoStop}%
\bibitem [{\citenamefont {Lee}\ \emph {et~al.}(2023)\citenamefont {Lee},
  \citenamefont {Kim}, \citenamefont {Bang}, \citenamefont {Park},
  \citenamefont {Kim}, \citenamefont {Wulferding},\ and\ \citenamefont
  {Cho}}]{Lee2023}%
  \BibitemOpen
  \bibfield  {author} {\bibinfo {author} {\bibfnamefont {S.}~\bibnamefont
  {Lee}}, \bibinfo {author} {\bibfnamefont {E.}~\bibnamefont {Kim}}, \bibinfo
  {author} {\bibfnamefont {J.}~\bibnamefont {Bang}}, \bibinfo {author}
  {\bibfnamefont {J.}~\bibnamefont {Park}}, \bibinfo {author} {\bibfnamefont
  {C.}~\bibnamefont {Kim}}, \bibinfo {author} {\bibfnamefont {D.}~\bibnamefont
  {Wulferding}},\ and\ \bibinfo {author} {\bibfnamefont {D.}~\bibnamefont
  {Cho}},\ }\bibfield  {title} {\bibinfo {title} {Melting of unidirectional
  charge density waves across twin domain boundaries in {GdTe$_3$}},\ }\href
  {https://doi.org/10.1021/acs.nanolett.3c03721} {\bibfield  {journal}
  {\bibinfo  {journal} {Nano Lett.}\ }\textbf {\bibinfo {volume} {23}},\
  \bibinfo {pages} {11219} (\bibinfo {year} {2023})}\BibitemShut {NoStop}%
\bibitem [{\citenamefont {Ru}\ \emph {et~al.}(2008)\citenamefont {Ru},
  \citenamefont {Chu},\ and\ \citenamefont {Fisher}}]{Ru2008}%
  \BibitemOpen
  \bibfield  {author} {\bibinfo {author} {\bibfnamefont {N.}~\bibnamefont
  {Ru}}, \bibinfo {author} {\bibfnamefont {J.-H.}\ \bibnamefont {Chu}},\ and\
  \bibinfo {author} {\bibfnamefont {I.~R.}\ \bibnamefont {Fisher}},\ }\bibfield
   {title} {\bibinfo {title} {Magnetic properties of the charge density wave
  compounds {$R{\text{Te}}_{3}$ ($R=\text{Y}$, La, Ce, Pr, Nd, Sm, Gd, Tb, Dy,
  Ho, Er, and Tm)}},\ }\href {https://doi.org/10.1103/PhysRevB.78.012410}
  {\bibfield  {journal} {\bibinfo  {journal} {Phys. Rev. B}\ }\textbf {\bibinfo
  {volume} {78}},\ \bibinfo {pages} {012410} (\bibinfo {year}
  {2008})}\BibitemShut {NoStop}%
\bibitem [{\citenamefont {Banerjee}\ \emph {et~al.}(2013)\citenamefont
  {Banerjee}, \citenamefont {Feng}, \citenamefont {Silevitch}, \citenamefont
  {Wang}, \citenamefont {Lang}, \citenamefont {Kuo}, \citenamefont {Fisher},\
  and\ \citenamefont {Rosenbaum}}]{Banerjee2013}%
  \BibitemOpen
  \bibfield  {author} {\bibinfo {author} {\bibfnamefont {A.}~\bibnamefont
  {Banerjee}}, \bibinfo {author} {\bibfnamefont {Y.}~\bibnamefont {Feng}},
  \bibinfo {author} {\bibfnamefont {D.~M.}\ \bibnamefont {Silevitch}}, \bibinfo
  {author} {\bibfnamefont {J.}~\bibnamefont {Wang}}, \bibinfo {author}
  {\bibfnamefont {J.~C.}\ \bibnamefont {Lang}}, \bibinfo {author}
  {\bibfnamefont {H.-H.}\ \bibnamefont {Kuo}}, \bibinfo {author} {\bibfnamefont
  {I.~R.}\ \bibnamefont {Fisher}},\ and\ \bibinfo {author} {\bibfnamefont
  {T.~F.}\ \bibnamefont {Rosenbaum}},\ }\bibfield  {title} {\bibinfo {title}
  {Charge transfer and multiple density waves in the rare earth tellurides},\
  }\href {https://doi.org/10.1103/PhysRevB.87.155131} {\bibfield  {journal}
  {\bibinfo  {journal} {Phys. Rev. B}\ }\textbf {\bibinfo {volume} {87}},\
  \bibinfo {pages} {155131} (\bibinfo {year} {2013})}\BibitemShut {NoStop}%
\bibitem [{\citenamefont {Liu}\ \emph {et~al.}(2020)\citenamefont {Liu},
  \citenamefont {Huan}, \citenamefont {Liu}, \citenamefont {Liu}, \citenamefont
  {Liu}, \citenamefont {Lu}, \citenamefont {Huang}, \citenamefont {Jiang},
  \citenamefont {Wang}, \citenamefont {Yu}, \citenamefont {Zou}, \citenamefont
  {Guo},\ and\ \citenamefont {Shen}}]{Liu2020}%
  \BibitemOpen
  \bibfield  {author} {\bibinfo {author} {\bibfnamefont {J.~S.}\ \bibnamefont
  {Liu}}, \bibinfo {author} {\bibfnamefont {S.~C.}\ \bibnamefont {Huan}},
  \bibinfo {author} {\bibfnamefont {Z.~H.}\ \bibnamefont {Liu}}, \bibinfo
  {author} {\bibfnamefont {W.~L.}\ \bibnamefont {Liu}}, \bibinfo {author}
  {\bibfnamefont {Z.~T.}\ \bibnamefont {Liu}}, \bibinfo {author} {\bibfnamefont
  {X.~L.}\ \bibnamefont {Lu}}, \bibinfo {author} {\bibfnamefont
  {Z.}~\bibnamefont {Huang}}, \bibinfo {author} {\bibfnamefont {Z.~C.}\
  \bibnamefont {Jiang}}, \bibinfo {author} {\bibfnamefont {X.}~\bibnamefont
  {Wang}}, \bibinfo {author} {\bibfnamefont {N.}~\bibnamefont {Yu}}, \bibinfo
  {author} {\bibfnamefont {Z.~Q.}\ \bibnamefont {Zou}}, \bibinfo {author}
  {\bibfnamefont {Y.~F.}\ \bibnamefont {Guo}},\ and\ \bibinfo {author}
  {\bibfnamefont {D.~W.}\ \bibnamefont {Shen}},\ }\bibfield  {title} {\bibinfo
  {title} {Electronic structure of the high-mobility two-dimensional
  antiferromagnetic metal {$\mathrm{Gd}{\mathrm{Te}}_{3}$}},\ }\href
  {https://doi.org/10.1103/PhysRevMaterials.4.114005} {\bibfield  {journal}
  {\bibinfo  {journal} {Phys. Rev. Mater.}\ }\textbf {\bibinfo {volume} {4}},\
  \bibinfo {pages} {114005} (\bibinfo {year} {2020})}\BibitemShut {NoStop}%
\bibitem [{\citenamefont {Wang}\ \emph {et~al.}(2022)\citenamefont {Wang},
  \citenamefont {Petrides}, \citenamefont {McNamara}, \citenamefont {Hosen},
  \citenamefont {Lei}, \citenamefont {Wu}, \citenamefont {Hart}, \citenamefont
  {Lv}, \citenamefont {Yan}, \citenamefont {Xiao}, \citenamefont {Cha},
  \citenamefont {Narang}, \citenamefont {Schoop},\ and\ \citenamefont
  {Burch}}]{Wang2022}%
  \BibitemOpen
  \bibfield  {author} {\bibinfo {author} {\bibfnamefont {Y.}~\bibnamefont
  {Wang}}, \bibinfo {author} {\bibfnamefont {I.}~\bibnamefont {Petrides}},
  \bibinfo {author} {\bibfnamefont {G.}~\bibnamefont {McNamara}}, \bibinfo
  {author} {\bibfnamefont {M.~M.}\ \bibnamefont {Hosen}}, \bibinfo {author}
  {\bibfnamefont {S.}~\bibnamefont {Lei}}, \bibinfo {author} {\bibfnamefont
  {Y.-C.}\ \bibnamefont {Wu}}, \bibinfo {author} {\bibfnamefont {J.~L.}\
  \bibnamefont {Hart}}, \bibinfo {author} {\bibfnamefont {H.}~\bibnamefont
  {Lv}}, \bibinfo {author} {\bibfnamefont {J.}~\bibnamefont {Yan}}, \bibinfo
  {author} {\bibfnamefont {D.}~\bibnamefont {Xiao}}, \bibinfo {author}
  {\bibfnamefont {J.~J.}\ \bibnamefont {Cha}}, \bibinfo {author} {\bibfnamefont
  {P.}~\bibnamefont {Narang}}, \bibinfo {author} {\bibfnamefont {L.~M.}\
  \bibnamefont {Schoop}},\ and\ \bibinfo {author} {\bibfnamefont {K.~S.}\
  \bibnamefont {Burch}},\ }\bibfield  {title} {\bibinfo {title} {Axial higgs
  mode detected by quantum pathway interference in {RTe$_3$}},\ }\href
  {https://doi.org/10.1038/s41586-022-04746-6} {\bibfield  {journal} {\bibinfo
  {journal} {Nature}\ }\textbf {\bibinfo {volume} {606}},\ \bibinfo {pages}
  {896} (\bibinfo {year} {2022})}\BibitemShut {NoStop}%
\bibitem [{\citenamefont {Rettig}\ \emph {et~al.}(2016)\citenamefont {Rettig},
  \citenamefont {Cort$\rm{\acute{e}}$s}, \citenamefont {Chu}, \citenamefont
  {Fisher}, \citenamefont {Schmitt}, \citenamefont {Moore}, \citenamefont
  {Shen}, \citenamefont {Kirchmann}, \citenamefont {Wolf},\ and\ \citenamefont
  {Bovensiepen}}]{Rettig2016}%
  \BibitemOpen
  \bibfield  {author} {\bibinfo {author} {\bibfnamefont {L.}~\bibnamefont
  {Rettig}}, \bibinfo {author} {\bibfnamefont {R.}~\bibnamefont
  {Cort$\rm{\acute{e}}$s}}, \bibinfo {author} {\bibfnamefont {J.-H.}\
  \bibnamefont {Chu}}, \bibinfo {author} {\bibfnamefont {I.~R.}\ \bibnamefont
  {Fisher}}, \bibinfo {author} {\bibfnamefont {F.}~\bibnamefont {Schmitt}},
  \bibinfo {author} {\bibfnamefont {R.~G.}\ \bibnamefont {Moore}}, \bibinfo
  {author} {\bibfnamefont {Z.-X.}\ \bibnamefont {Shen}}, \bibinfo {author}
  {\bibfnamefont {P.~S.}\ \bibnamefont {Kirchmann}}, \bibinfo {author}
  {\bibfnamefont {M.}~\bibnamefont {Wolf}},\ and\ \bibinfo {author}
  {\bibfnamefont {U.}~\bibnamefont {Bovensiepen}},\ }\bibfield  {title}
  {\bibinfo {title} {Persistent order due to transiently enhanced nesting in an
  electronically excited charge density wave},\ }\href
  {https://doi.org/10.1038/ncomms10459} {\bibfield  {journal} {\bibinfo
  {journal} {Nat. Commun.}\ }\textbf {\bibinfo {volume} {7}},\ \bibinfo {pages}
  {10459} (\bibinfo {year} {2016})}\BibitemShut {NoStop}%
\bibitem [{\citenamefont {Gonzalez-Vallejo}\ \emph {et~al.}(2022)\citenamefont
  {Gonzalez-Vallejo}, \citenamefont {Jacques}, \citenamefont {Boschetto},
  \citenamefont {Rizza}, \citenamefont {Hadj-Azzem}, \citenamefont {Faure},\
  and\ \citenamefont {Le~Bolloc'h}}]{GonzalezVallejo2022}%
  \BibitemOpen
  \bibfield  {author} {\bibinfo {author} {\bibfnamefont {I.}~\bibnamefont
  {Gonzalez-Vallejo}}, \bibinfo {author} {\bibfnamefont {V.~L.~R.}\
  \bibnamefont {Jacques}}, \bibinfo {author} {\bibfnamefont {D.}~\bibnamefont
  {Boschetto}}, \bibinfo {author} {\bibfnamefont {G.}~\bibnamefont {Rizza}},
  \bibinfo {author} {\bibfnamefont {A.}~\bibnamefont {Hadj-Azzem}}, \bibinfo
  {author} {\bibfnamefont {J.}~\bibnamefont {Faure}},\ and\ \bibinfo {author}
  {\bibfnamefont {D.}~\bibnamefont {Le~Bolloc'h}},\ }\bibfield  {title}
  {\bibinfo {title} {{Time-resolved structural dynamics of the
  out-of-equilibrium charge density wave phase transition in {GdTe$_3$}}},\
  }\href {https://doi.org/10.1063/4.0000131} {\bibfield  {journal} {\bibinfo
  {journal} {Struct. Dynam.}\ }\textbf {\bibinfo {volume} {9}},\ \bibinfo
  {pages} {014502} (\bibinfo {year} {2022})}\BibitemShut {NoStop}%
\bibitem [{\citenamefont {Zong}\ \emph
  {et~al.}(2019{\natexlab{a}})\citenamefont {Zong}, \citenamefont {Kogar},
  \citenamefont {Bie}, \citenamefont {Rohwer}, \citenamefont {Lee},
  \citenamefont {Baldini}, \citenamefont {Ergecen}, \citenamefont {Yilmaz},
  \citenamefont {Freelon}, \citenamefont {Sie}, \citenamefont {Zhou},
  \citenamefont {Straquadine}, \citenamefont {Walmsley}, \citenamefont
  {Dolgirev}, \citenamefont {Rozhkov}, \citenamefont {Fisher}, \citenamefont
  {Jarillo-Herrero}, \citenamefont {Fine},\ and\ \citenamefont
  {Gedik}}]{Zong2019}%
  \BibitemOpen
  \bibfield  {author} {\bibinfo {author} {\bibfnamefont {A.}~\bibnamefont
  {Zong}}, \bibinfo {author} {\bibfnamefont {A.}~\bibnamefont {Kogar}},
  \bibinfo {author} {\bibfnamefont {Y.-Q.}\ \bibnamefont {Bie}}, \bibinfo
  {author} {\bibfnamefont {T.}~\bibnamefont {Rohwer}}, \bibinfo {author}
  {\bibfnamefont {C.}~\bibnamefont {Lee}}, \bibinfo {author} {\bibfnamefont
  {E.}~\bibnamefont {Baldini}}, \bibinfo {author} {\bibfnamefont
  {E.}~\bibnamefont {Ergecen}}, \bibinfo {author} {\bibfnamefont {M.~B.}\
  \bibnamefont {Yilmaz}}, \bibinfo {author} {\bibfnamefont {B.}~\bibnamefont
  {Freelon}}, \bibinfo {author} {\bibfnamefont {E.~J.}\ \bibnamefont {Sie}},
  \bibinfo {author} {\bibfnamefont {H.}~\bibnamefont {Zhou}}, \bibinfo {author}
  {\bibfnamefont {J.}~\bibnamefont {Straquadine}}, \bibinfo {author}
  {\bibfnamefont {P.}~\bibnamefont {Walmsley}}, \bibinfo {author}
  {\bibfnamefont {P.~E.}\ \bibnamefont {Dolgirev}}, \bibinfo {author}
  {\bibfnamefont {A.~V.}\ \bibnamefont {Rozhkov}}, \bibinfo {author}
  {\bibfnamefont {I.~R.}\ \bibnamefont {Fisher}}, \bibinfo {author}
  {\bibfnamefont {P.}~\bibnamefont {Jarillo-Herrero}}, \bibinfo {author}
  {\bibfnamefont {B.~V.}\ \bibnamefont {Fine}},\ and\ \bibinfo {author}
  {\bibfnamefont {N.}~\bibnamefont {Gedik}},\ }\bibfield  {title} {\bibinfo
  {title} {Evidence for topological defects in a photoinduced phase
  transition},\ }\href {https://doi.org/10.1038/s41567-018-0311-9} {\bibfield
  {journal} {\bibinfo  {journal} {Nat. Phys.}\ }\textbf {\bibinfo {volume}
  {15}},\ \bibinfo {pages} {27} (\bibinfo {year}
  {2019}{\natexlab{a}})}\BibitemShut {NoStop}%
\bibitem [{\citenamefont {Zong}\ \emph
  {et~al.}(2019{\natexlab{b}})\citenamefont {Zong}, \citenamefont {Dolgirev},
  \citenamefont {Kogar}, \citenamefont {Erge\ifmmode~\mbox{\c{c}}\else
  \c{c}\fi{}en}, \citenamefont {Yilmaz}, \citenamefont {Bie}, \citenamefont
  {Rohwer}, \citenamefont {Tung}, \citenamefont {Straquadine}, \citenamefont
  {Wang}, \citenamefont {Yang}, \citenamefont {Shen}, \citenamefont {Li},
  \citenamefont {Yang}, \citenamefont {Park}, \citenamefont {Hoffmann},
  \citenamefont {Ofori-Okai}, \citenamefont {Kozina}, \citenamefont {Wen},
  \citenamefont {Wang}, \citenamefont {Fisher}, \citenamefont
  {Jarillo-Herrero},\ and\ \citenamefont {Gedik}}]{Zong2019a}%
  \BibitemOpen
  \bibfield  {author} {\bibinfo {author} {\bibfnamefont {A.}~\bibnamefont
  {Zong}}, \bibinfo {author} {\bibfnamefont {P.~E.}\ \bibnamefont {Dolgirev}},
  \bibinfo {author} {\bibfnamefont {A.}~\bibnamefont {Kogar}}, \bibinfo
  {author} {\bibfnamefont {E.}~\bibnamefont {Erge\ifmmode~\mbox{\c{c}}\else
  \c{c}\fi{}en}}, \bibinfo {author} {\bibfnamefont {M.~B.}\ \bibnamefont
  {Yilmaz}}, \bibinfo {author} {\bibfnamefont {Y.-Q.}\ \bibnamefont {Bie}},
  \bibinfo {author} {\bibfnamefont {T.}~\bibnamefont {Rohwer}}, \bibinfo
  {author} {\bibfnamefont {I.-C.}\ \bibnamefont {Tung}}, \bibinfo {author}
  {\bibfnamefont {J.}~\bibnamefont {Straquadine}}, \bibinfo {author}
  {\bibfnamefont {X.}~\bibnamefont {Wang}}, \bibinfo {author} {\bibfnamefont
  {Y.}~\bibnamefont {Yang}}, \bibinfo {author} {\bibfnamefont {X.}~\bibnamefont
  {Shen}}, \bibinfo {author} {\bibfnamefont {R.}~\bibnamefont {Li}}, \bibinfo
  {author} {\bibfnamefont {J.}~\bibnamefont {Yang}}, \bibinfo {author}
  {\bibfnamefont {S.}~\bibnamefont {Park}}, \bibinfo {author} {\bibfnamefont
  {M.~C.}\ \bibnamefont {Hoffmann}}, \bibinfo {author} {\bibfnamefont {B.~K.}\
  \bibnamefont {Ofori-Okai}}, \bibinfo {author} {\bibfnamefont {M.~E.}\
  \bibnamefont {Kozina}}, \bibinfo {author} {\bibfnamefont {H.}~\bibnamefont
  {Wen}}, \bibinfo {author} {\bibfnamefont {X.}~\bibnamefont {Wang}}, \bibinfo
  {author} {\bibfnamefont {I.~R.}\ \bibnamefont {Fisher}}, \bibinfo {author}
  {\bibfnamefont {P.}~\bibnamefont {Jarillo-Herrero}},\ and\ \bibinfo {author}
  {\bibfnamefont {N.}~\bibnamefont {Gedik}},\ }\bibfield  {title} {\bibinfo
  {title} {Dynamical slowing-down in an ultrafast photoinduced phase
  transition},\ }\href {https://doi.org/10.1103/PhysRevLett.123.097601}
  {\bibfield  {journal} {\bibinfo  {journal} {Phys. Rev. Lett.}\ }\textbf
  {\bibinfo {volume} {123}},\ \bibinfo {pages} {097601} (\bibinfo {year}
  {2019}{\natexlab{b}})}\BibitemShut {NoStop}%
\bibitem [{\citenamefont {Gon\c{c}alves-Faria}\ \emph
  {et~al.}(2024)\citenamefont {Gon\c{c}alves-Faria}, \citenamefont {Pashkin},
  \citenamefont {Wang}, \citenamefont {Lei}, \citenamefont {Winnerl},
  \citenamefont {Tsirlin}, \citenamefont {Helm},\ and\ \citenamefont
  {Uykur}}]{GoncalvesFaria2024}%
  \BibitemOpen
  \bibfield  {author} {\bibinfo {author} {\bibfnamefont {M.~V.}\ \bibnamefont
  {Gon\c{c}alves-Faria}}, \bibinfo {author} {\bibfnamefont {A.}~\bibnamefont
  {Pashkin}}, \bibinfo {author} {\bibfnamefont {Q.}~\bibnamefont {Wang}},
  \bibinfo {author} {\bibfnamefont {H.~C.}\ \bibnamefont {Lei}}, \bibinfo
  {author} {\bibfnamefont {S.}~\bibnamefont {Winnerl}}, \bibinfo {author}
  {\bibfnamefont {A.~A.}\ \bibnamefont {Tsirlin}}, \bibinfo {author}
  {\bibfnamefont {M.}~\bibnamefont {Helm}},\ and\ \bibinfo {author}
  {\bibfnamefont {E.}~\bibnamefont {Uykur}},\ }\bibfield  {title} {\bibinfo
  {title} {Coherent phonon and unconventional carriers in the magnetic kagome
  metal {Fe$_3$Sn$_2$}},\ }\href {https://doi.org/10.1038/s41535-024-00642-6}
  {\bibfield  {journal} {\bibinfo  {journal} {npj Quantum Materials}\ }\textbf
  {\bibinfo {volume} {9}},\ \bibinfo {pages} {31} (\bibinfo {year}
  {2024})}\BibitemShut {NoStop}%
\bibitem [{\citenamefont {Leuenberger}\ \emph {et~al.}(2015)\citenamefont
  {Leuenberger}, \citenamefont {Sobota}, \citenamefont {Yang}, \citenamefont
  {Kemper}, \citenamefont {Giraldo-Gallo}, \citenamefont {Moore}, \citenamefont
  {Fisher}, \citenamefont {Kirchmann}, \citenamefont {Devereaux},\ and\
  \citenamefont {Shen}}]{Leuenberger2015}%
  \BibitemOpen
  \bibfield  {author} {\bibinfo {author} {\bibfnamefont {D.}~\bibnamefont
  {Leuenberger}}, \bibinfo {author} {\bibfnamefont {J.~A.}\ \bibnamefont
  {Sobota}}, \bibinfo {author} {\bibfnamefont {S.-L.}\ \bibnamefont {Yang}},
  \bibinfo {author} {\bibfnamefont {A.~F.}\ \bibnamefont {Kemper}}, \bibinfo
  {author} {\bibfnamefont {P.}~\bibnamefont {Giraldo-Gallo}}, \bibinfo {author}
  {\bibfnamefont {R.~G.}\ \bibnamefont {Moore}}, \bibinfo {author}
  {\bibfnamefont {I.~R.}\ \bibnamefont {Fisher}}, \bibinfo {author}
  {\bibfnamefont {P.~S.}\ \bibnamefont {Kirchmann}}, \bibinfo {author}
  {\bibfnamefont {T.~P.}\ \bibnamefont {Devereaux}},\ and\ \bibinfo {author}
  {\bibfnamefont {Z.-X.}\ \bibnamefont {Shen}},\ }\bibfield  {title} {\bibinfo
  {title} {Classification of collective modes in a charge density wave by
  momentum-dependent modulation of the electronic band structure},\ }\href
  {https://doi.org/10.1103/PhysRevB.91.201106} {\bibfield  {journal} {\bibinfo
  {journal} {Phys. Rev. B}\ }\textbf {\bibinfo {volume} {91}},\ \bibinfo
  {pages} {201106} (\bibinfo {year} {2015})}\BibitemShut {NoStop}%
\bibitem [{\citenamefont {Schmitt}\ \emph {et~al.}(2008)\citenamefont
  {Schmitt}, \citenamefont {Kirchmann}, \citenamefont {Bovensiepen},
  \citenamefont {Moore}, \citenamefont {Rettig}, \citenamefont {Krenz},
  \citenamefont {Chu}, \citenamefont {Ru}, \citenamefont {Perfetti},
  \citenamefont {Lu}, \citenamefont {Wolf}, \citenamefont {Fisher},\ and\
  \citenamefont {Shen}}]{Schmitt2008}%
  \BibitemOpen
  \bibfield  {author} {\bibinfo {author} {\bibfnamefont {F.}~\bibnamefont
  {Schmitt}}, \bibinfo {author} {\bibfnamefont {P.~S.}\ \bibnamefont
  {Kirchmann}}, \bibinfo {author} {\bibfnamefont {U.}~\bibnamefont
  {Bovensiepen}}, \bibinfo {author} {\bibfnamefont {R.~G.}\ \bibnamefont
  {Moore}}, \bibinfo {author} {\bibfnamefont {L.}~\bibnamefont {Rettig}},
  \bibinfo {author} {\bibfnamefont {M.}~\bibnamefont {Krenz}}, \bibinfo
  {author} {\bibfnamefont {J.-H.}\ \bibnamefont {Chu}}, \bibinfo {author}
  {\bibfnamefont {N.}~\bibnamefont {Ru}}, \bibinfo {author} {\bibfnamefont
  {L.}~\bibnamefont {Perfetti}}, \bibinfo {author} {\bibfnamefont {D.~H.}\
  \bibnamefont {Lu}}, \bibinfo {author} {\bibfnamefont {M.}~\bibnamefont
  {Wolf}}, \bibinfo {author} {\bibfnamefont {I.~R.}\ \bibnamefont {Fisher}},\
  and\ \bibinfo {author} {\bibfnamefont {Z.-X.}\ \bibnamefont {Shen}},\
  }\bibfield  {title} {\bibinfo {title} {Transient electronic structure and
  melting of a charge density wave in {TbTe$_3$}},\ }\href
  {https://doi.org/10.1126/science.1160778} {\bibfield  {journal} {\bibinfo
  {journal} {Science}\ }\textbf {\bibinfo {volume} {321}},\ \bibinfo {pages}
  {1649} (\bibinfo {year} {2008})}\BibitemShut {NoStop}%
\bibitem [{\citenamefont {Rettig}\ \emph {et~al.}(2014)\citenamefont {Rettig},
  \citenamefont {Chu}, \citenamefont {Fisher}, \citenamefont {Bovensiepen},\
  and\ \citenamefont {Wolf}}]{Rettig2014}%
  \BibitemOpen
  \bibfield  {author} {\bibinfo {author} {\bibfnamefont {L.}~\bibnamefont
  {Rettig}}, \bibinfo {author} {\bibfnamefont {J.-H.}\ \bibnamefont {Chu}},
  \bibinfo {author} {\bibfnamefont {I.~R.}\ \bibnamefont {Fisher}}, \bibinfo
  {author} {\bibfnamefont {U.}~\bibnamefont {Bovensiepen}},\ and\ \bibinfo
  {author} {\bibfnamefont {M.}~\bibnamefont {Wolf}},\ }\bibfield  {title}
  {\bibinfo {title} {Coherent dynamics of the charge density wave gap in
  tritellurides},\ }\href {https://doi.org/10.1039/C4FD00045E} {\bibfield
  {journal} {\bibinfo  {journal} {Faraday Discuss.}\ }\textbf {\bibinfo
  {volume} {171}},\ \bibinfo {pages} {299} (\bibinfo {year}
  {2014})}\BibitemShut {NoStop}%
\bibitem [{\citenamefont {Dutta}\ \emph {et~al.}(2024)\citenamefont {Dutta},
  \citenamefont {Chandra}, \citenamefont {Maria}, \citenamefont {Debnath},
  \citenamefont {Rawat}, \citenamefont {Soni}, \citenamefont {Waghmare},\ and\
  \citenamefont {Biswas}}]{Dutta}%
  \BibitemOpen
  \bibfield  {author} {\bibinfo {author} {\bibfnamefont {P.}~\bibnamefont
  {Dutta}}, \bibinfo {author} {\bibfnamefont {S.}~\bibnamefont {Chandra}},
  \bibinfo {author} {\bibfnamefont {I.}~\bibnamefont {Maria}}, \bibinfo
  {author} {\bibfnamefont {K.}~\bibnamefont {Debnath}}, \bibinfo {author}
  {\bibfnamefont {D.}~\bibnamefont {Rawat}}, \bibinfo {author} {\bibfnamefont
  {A.}~\bibnamefont {Soni}}, \bibinfo {author} {\bibfnamefont {U.~V.}\
  \bibnamefont {Waghmare}},\ and\ \bibinfo {author} {\bibfnamefont
  {K.}~\bibnamefont {Biswas}},\ }\bibfield  {title} {\bibinfo {title} {Lattice
  instability induced concerted structural distortion in charged and van der
  waals layered {GdTe$_3$}},\ }\href
  {https://doi.org/https://doi.org/10.1002/adfm.202312663} {\bibfield
  {journal} {\bibinfo  {journal} {Adv. Funct. Mater.}\ }\textbf {\bibinfo
  {volume} {34}},\ \bibinfo {pages} {2312663} (\bibinfo {year}
  {2024})}\BibitemShut {NoStop}%
\bibitem [{\citenamefont {Lavagnini}\ \emph {et~al.}(2010)\citenamefont
  {Lavagnini}, \citenamefont {Eiter}, \citenamefont {Tassini}, \citenamefont
  {Muschler}, \citenamefont {Hackl}, \citenamefont {Monnier}, \citenamefont
  {Chu}, \citenamefont {Fisher},\ and\ \citenamefont
  {Degiorgi}}]{Lavagnini2010}%
  \BibitemOpen
  \bibfield  {author} {\bibinfo {author} {\bibfnamefont {M.}~\bibnamefont
  {Lavagnini}}, \bibinfo {author} {\bibfnamefont {H.-M.}\ \bibnamefont
  {Eiter}}, \bibinfo {author} {\bibfnamefont {L.}~\bibnamefont {Tassini}},
  \bibinfo {author} {\bibfnamefont {B.}~\bibnamefont {Muschler}}, \bibinfo
  {author} {\bibfnamefont {R.}~\bibnamefont {Hackl}}, \bibinfo {author}
  {\bibfnamefont {R.}~\bibnamefont {Monnier}}, \bibinfo {author} {\bibfnamefont
  {J.-H.}\ \bibnamefont {Chu}}, \bibinfo {author} {\bibfnamefont {I.~R.}\
  \bibnamefont {Fisher}},\ and\ \bibinfo {author} {\bibfnamefont
  {L.}~\bibnamefont {Degiorgi}},\ }\bibfield  {title} {\bibinfo {title} {Raman
  scattering evidence for a cascade evolution of the charge-density-wave
  collective amplitude mode},\ }\href
  {https://doi.org/10.1103/PhysRevB.81.081101} {\bibfield  {journal} {\bibinfo
  {journal} {Phys. Rev. B}\ }\textbf {\bibinfo {volume} {81}},\ \bibinfo
  {pages} {081101} (\bibinfo {year} {2010})}\BibitemShut {NoStop}%
\bibitem [{\citenamefont {Yusupov}\ \emph {et~al.}(2008)\citenamefont
  {Yusupov}, \citenamefont {Mertelj}, \citenamefont {Chu}, \citenamefont
  {Fisher},\ and\ \citenamefont {Mihailovic}}]{Yusupov2008}%
  \BibitemOpen
  \bibfield  {author} {\bibinfo {author} {\bibfnamefont {R.~V.}\ \bibnamefont
  {Yusupov}}, \bibinfo {author} {\bibfnamefont {T.}~\bibnamefont {Mertelj}},
  \bibinfo {author} {\bibfnamefont {J.-H.}\ \bibnamefont {Chu}}, \bibinfo
  {author} {\bibfnamefont {I.~R.}\ \bibnamefont {Fisher}},\ and\ \bibinfo
  {author} {\bibfnamefont {D.}~\bibnamefont {Mihailovic}},\ }\bibfield  {title}
  {\bibinfo {title} {Single-particle and collective mode couplings associated
  with 1- and 2-directional electronic ordering in metallic
  {$R{\mathrm{Te}}_{3}$ ($R=\mathrm{Ho},\mathrm{Dy},\mathrm{Tb}$)}},\ }\href
  {https://doi.org/10.1103/PhysRevLett.101.246402} {\bibfield  {journal}
  {\bibinfo  {journal} {Phys. Rev. Lett.}\ }\textbf {\bibinfo {volume} {101}},\
  \bibinfo {pages} {246402} (\bibinfo {year} {2008})}\BibitemShut {NoStop}%
\bibitem [{\citenamefont {Moore}\ \emph {et~al.}(2016)\citenamefont {Moore},
  \citenamefont {Lee}, \citenamefont {Kirchman}, \citenamefont {Chuang},
  \citenamefont {Kemper}, \citenamefont {Trigo}, \citenamefont {Patthey},
  \citenamefont {Lu}, \citenamefont {Krupin}, \citenamefont {Yi}, \citenamefont
  {Reis}, \citenamefont {Doering}, \citenamefont {Denes}, \citenamefont
  {Schlotter}, \citenamefont {Turner}, \citenamefont {Hays}, \citenamefont
  {Hering}, \citenamefont {Benson}, \citenamefont {Chu}, \citenamefont
  {Devereaux}, \citenamefont {Fisher}, \citenamefont {Hussain},\ and\
  \citenamefont {Shen}}]{Moore2016}%
  \BibitemOpen
  \bibfield  {author} {\bibinfo {author} {\bibfnamefont {R.~G.}\ \bibnamefont
  {Moore}}, \bibinfo {author} {\bibfnamefont {W.~S.}\ \bibnamefont {Lee}},
  \bibinfo {author} {\bibfnamefont {P.~S.}\ \bibnamefont {Kirchman}}, \bibinfo
  {author} {\bibfnamefont {Y.~D.}\ \bibnamefont {Chuang}}, \bibinfo {author}
  {\bibfnamefont {A.~F.}\ \bibnamefont {Kemper}}, \bibinfo {author}
  {\bibfnamefont {M.}~\bibnamefont {Trigo}}, \bibinfo {author} {\bibfnamefont
  {L.}~\bibnamefont {Patthey}}, \bibinfo {author} {\bibfnamefont {D.~H.}\
  \bibnamefont {Lu}}, \bibinfo {author} {\bibfnamefont {O.}~\bibnamefont
  {Krupin}}, \bibinfo {author} {\bibfnamefont {M.}~\bibnamefont {Yi}}, \bibinfo
  {author} {\bibfnamefont {D.~A.}\ \bibnamefont {Reis}}, \bibinfo {author}
  {\bibfnamefont {D.}~\bibnamefont {Doering}}, \bibinfo {author} {\bibfnamefont
  {P.}~\bibnamefont {Denes}}, \bibinfo {author} {\bibfnamefont {W.~F.}\
  \bibnamefont {Schlotter}}, \bibinfo {author} {\bibfnamefont {J.~J.}\
  \bibnamefont {Turner}}, \bibinfo {author} {\bibfnamefont {G.}~\bibnamefont
  {Hays}}, \bibinfo {author} {\bibfnamefont {P.}~\bibnamefont {Hering}},
  \bibinfo {author} {\bibfnamefont {T.}~\bibnamefont {Benson}}, \bibinfo
  {author} {\bibfnamefont {J.-H.}\ \bibnamefont {Chu}}, \bibinfo {author}
  {\bibfnamefont {T.~P.}\ \bibnamefont {Devereaux}}, \bibinfo {author}
  {\bibfnamefont {I.~R.}\ \bibnamefont {Fisher}}, \bibinfo {author}
  {\bibfnamefont {Z.}~\bibnamefont {Hussain}},\ and\ \bibinfo {author}
  {\bibfnamefont {Z.-X.}\ \bibnamefont {Shen}},\ }\bibfield  {title} {\bibinfo
  {title} {Ultrafast resonant soft x-ray diffraction dynamics of the charge
  density wave in {${\mathrm{TbTe}}_{3}$}},\ }\href
  {https://doi.org/10.1103/PhysRevB.93.024304} {\bibfield  {journal} {\bibinfo
  {journal} {Phys. Rev. B}\ }\textbf {\bibinfo {volume} {93}},\ \bibinfo
  {pages} {024304} (\bibinfo {year} {2016})}\BibitemShut {NoStop}%
\bibitem [{\citenamefont {Boschini}\ \emph {et~al.}(2020)\citenamefont
  {Boschini}, \citenamefont {Bugini}, \citenamefont {Zonno}, \citenamefont
  {Michiardi}, \citenamefont {Day}, \citenamefont {Razzoli}, \citenamefont
  {Zwartsenberg}, \citenamefont {Schneider}, \citenamefont {da~Silva~Neto},
  \citenamefont {dal Conte}, \citenamefont {Kushwaha}, \citenamefont {Cava},
  \citenamefont {Zhdanovich}, \citenamefont {Mills}, \citenamefont {Levy},
  \citenamefont {Carpene}, \citenamefont {Dallera}, \citenamefont {Giannetti},
  \citenamefont {Jones}, \citenamefont {Cerullo},\ and\ \citenamefont
  {Damascelli}}]{Boschini2020}%
  \BibitemOpen
  \bibfield  {author} {\bibinfo {author} {\bibfnamefont {F.}~\bibnamefont
  {Boschini}}, \bibinfo {author} {\bibfnamefont {D.}~\bibnamefont {Bugini}},
  \bibinfo {author} {\bibfnamefont {M.}~\bibnamefont {Zonno}}, \bibinfo
  {author} {\bibfnamefont {M.}~\bibnamefont {Michiardi}}, \bibinfo {author}
  {\bibfnamefont {R.~P.}\ \bibnamefont {Day}}, \bibinfo {author} {\bibfnamefont
  {E.}~\bibnamefont {Razzoli}}, \bibinfo {author} {\bibfnamefont
  {B.}~\bibnamefont {Zwartsenberg}}, \bibinfo {author} {\bibfnamefont
  {M.}~\bibnamefont {Schneider}}, \bibinfo {author} {\bibfnamefont {E.~H.}\
  \bibnamefont {da~Silva~Neto}}, \bibinfo {author} {\bibfnamefont
  {S.}~\bibnamefont {dal Conte}}, \bibinfo {author} {\bibfnamefont {S.~K.}\
  \bibnamefont {Kushwaha}}, \bibinfo {author} {\bibfnamefont {R.~J.}\
  \bibnamefont {Cava}}, \bibinfo {author} {\bibfnamefont {S.}~\bibnamefont
  {Zhdanovich}}, \bibinfo {author} {\bibfnamefont {A.~K.}\ \bibnamefont
  {Mills}}, \bibinfo {author} {\bibfnamefont {G.}~\bibnamefont {Levy}},
  \bibinfo {author} {\bibfnamefont {E.}~\bibnamefont {Carpene}}, \bibinfo
  {author} {\bibfnamefont {C.}~\bibnamefont {Dallera}}, \bibinfo {author}
  {\bibfnamefont {C.}~\bibnamefont {Giannetti}}, \bibinfo {author}
  {\bibfnamefont {D.~J.}\ \bibnamefont {Jones}}, \bibinfo {author}
  {\bibfnamefont {G.}~\bibnamefont {Cerullo}},\ and\ \bibinfo {author}
  {\bibfnamefont {A.}~\bibnamefont {Damascelli}},\ }\bibfield  {title}
  {\bibinfo {title} {Role of matrix elements in the time-resolved photoemission
  signal},\ }\href {https://doi.org/10.1088/1367-2630/ab6eb1} {\bibfield
  {journal} {\bibinfo  {journal} {New Journal of Physics}\ }\textbf {\bibinfo
  {volume} {22}},\ \bibinfo {pages} {023031} (\bibinfo {year}
  {2020})}\BibitemShut {NoStop}%
\bibitem [{\citenamefont {Seah}\ and\ \citenamefont {Dench}(1979)}]{Seah1979}%
  \BibitemOpen
  \bibfield  {author} {\bibinfo {author} {\bibfnamefont {M.~P.}\ \bibnamefont
  {Seah}}\ and\ \bibinfo {author} {\bibfnamefont {W.~A.}\ \bibnamefont
  {Dench}},\ }\bibfield  {title} {\bibinfo {title} {Quantitative electron
  spectroscopy of surfaces: A standard data base for electron inelastic mean
  free paths in solids},\ }\href {https://doi.org/10.1002/sia.740010103}
  {\bibfield  {journal} {\bibinfo  {journal} {Surf. Interface Anal.}\ }\textbf
  {\bibinfo {volume} {1}},\ \bibinfo {pages} {2} (\bibinfo {year}
  {1979})}\BibitemShut {NoStop}%
\bibitem [{\citenamefont {Clarke}\ \emph {et~al.}(1994)\citenamefont {Clarke},
  \citenamefont {Hastie}, \citenamefont {Kihlborg}, \citenamefont {Metselaar},\
  and\ \citenamefont {Thackeray}}]{Clarke1994}%
  \BibitemOpen
  \bibfield  {author} {\bibinfo {author} {\bibfnamefont {J.~B.}\ \bibnamefont
  {Clarke}}, \bibinfo {author} {\bibfnamefont {J.~W.}\ \bibnamefont {Hastie}},
  \bibinfo {author} {\bibfnamefont {L.~H.~E.}\ \bibnamefont {Kihlborg}},
  \bibinfo {author} {\bibfnamefont {R.}~\bibnamefont {Metselaar}},\ and\
  \bibinfo {author} {\bibfnamefont {M.~M.}\ \bibnamefont {Thackeray}},\
  }\bibfield  {title} {\bibinfo {title} {Definitions of terms relating to phase
  transitions of the solid state },\ }\href
  {https://doi.org/doi:10.1351/pac199466030577} {\bibfield  {journal} {\bibinfo
   {journal} {Pure Appl. Chem.}\ }\textbf {\bibinfo {volume} {66}},\ \bibinfo
  {pages} {577} (\bibinfo {year} {1994})}\BibitemShut {NoStop}%
\bibitem [{\citenamefont {Trigo}\ \emph {et~al.}(2019)\citenamefont {Trigo},
  \citenamefont {Giraldo-Gallo}, \citenamefont {Kozina}, \citenamefont
  {Henighan}, \citenamefont {Jiang}, \citenamefont {Liu}, \citenamefont
  {Clark}, \citenamefont {Chollet}, \citenamefont {Glownia}, \citenamefont
  {Zhu}, \citenamefont {Katayama}, \citenamefont {Leuenberger}, \citenamefont
  {Kirchmann}, \citenamefont {Fisher}, \citenamefont {Shen},\ and\
  \citenamefont {Reis}}]{Trigo2019}%
  \BibitemOpen
  \bibfield  {author} {\bibinfo {author} {\bibfnamefont {M.}~\bibnamefont
  {Trigo}}, \bibinfo {author} {\bibfnamefont {P.}~\bibnamefont
  {Giraldo-Gallo}}, \bibinfo {author} {\bibfnamefont {M.~E.}\ \bibnamefont
  {Kozina}}, \bibinfo {author} {\bibfnamefont {T.}~\bibnamefont {Henighan}},
  \bibinfo {author} {\bibfnamefont {M.~P.}\ \bibnamefont {Jiang}}, \bibinfo
  {author} {\bibfnamefont {H.}~\bibnamefont {Liu}}, \bibinfo {author}
  {\bibfnamefont {J.~N.}\ \bibnamefont {Clark}}, \bibinfo {author}
  {\bibfnamefont {M.}~\bibnamefont {Chollet}}, \bibinfo {author} {\bibfnamefont
  {J.~M.}\ \bibnamefont {Glownia}}, \bibinfo {author} {\bibfnamefont
  {D.}~\bibnamefont {Zhu}}, \bibinfo {author} {\bibfnamefont {T.}~\bibnamefont
  {Katayama}}, \bibinfo {author} {\bibfnamefont {D.}~\bibnamefont
  {Leuenberger}}, \bibinfo {author} {\bibfnamefont {P.~S.}\ \bibnamefont
  {Kirchmann}}, \bibinfo {author} {\bibfnamefont {I.~R.}\ \bibnamefont
  {Fisher}}, \bibinfo {author} {\bibfnamefont {Z.~X.}\ \bibnamefont {Shen}},\
  and\ \bibinfo {author} {\bibfnamefont {D.~A.}\ \bibnamefont {Reis}},\
  }\bibfield  {title} {\bibinfo {title} {Coherent order parameter dynamics in
  {${\mathrm{SmTe}}_{3}$}},\ }\href
  {https://doi.org/10.1103/PhysRevB.99.104111} {\bibfield  {journal} {\bibinfo
  {journal} {Phys. Rev. B}\ }\textbf {\bibinfo {volume} {99}},\ \bibinfo
  {pages} {104111} (\bibinfo {year} {2019})}\BibitemShut {NoStop}%
\bibitem [{\citenamefont {Huber}\ \emph {et~al.}(2014)\citenamefont {Huber},
  \citenamefont {Mariager}, \citenamefont {Ferrer}, \citenamefont {Sch\"afer},
  \citenamefont {Johnson}, \citenamefont {Gr\"ubel}, \citenamefont {L\"ubcke},
  \citenamefont {Huber}, \citenamefont {Kubacka}, \citenamefont {Dornes},
  \citenamefont {Laulhe}, \citenamefont {Ravy}, \citenamefont {Ingold},
  \citenamefont {Beaud}, \citenamefont {Demsar},\ and\ \citenamefont
  {Johnson}}]{Huber2014}%
  \BibitemOpen
  \bibfield  {author} {\bibinfo {author} {\bibfnamefont {T.}~\bibnamefont
  {Huber}}, \bibinfo {author} {\bibfnamefont {S.~O.}\ \bibnamefont {Mariager}},
  \bibinfo {author} {\bibfnamefont {A.}~\bibnamefont {Ferrer}}, \bibinfo
  {author} {\bibfnamefont {H.}~\bibnamefont {Sch\"afer}}, \bibinfo {author}
  {\bibfnamefont {J.~A.}\ \bibnamefont {Johnson}}, \bibinfo {author}
  {\bibfnamefont {S.}~\bibnamefont {Gr\"ubel}}, \bibinfo {author}
  {\bibfnamefont {A.}~\bibnamefont {L\"ubcke}}, \bibinfo {author}
  {\bibfnamefont {L.}~\bibnamefont {Huber}}, \bibinfo {author} {\bibfnamefont
  {T.}~\bibnamefont {Kubacka}}, \bibinfo {author} {\bibfnamefont
  {C.}~\bibnamefont {Dornes}}, \bibinfo {author} {\bibfnamefont
  {C.}~\bibnamefont {Laulhe}}, \bibinfo {author} {\bibfnamefont
  {S.}~\bibnamefont {Ravy}}, \bibinfo {author} {\bibfnamefont {G.}~\bibnamefont
  {Ingold}}, \bibinfo {author} {\bibfnamefont {P.}~\bibnamefont {Beaud}},
  \bibinfo {author} {\bibfnamefont {J.}~\bibnamefont {Demsar}},\ and\ \bibinfo
  {author} {\bibfnamefont {S.~L.}\ \bibnamefont {Johnson}},\ }\bibfield
  {title} {\bibinfo {title} {Coherent structural dynamics of a prototypical
  charge-density-wave-to-metal transition},\ }\href
  {https://doi.org/10.1103/PhysRevLett.113.026401} {\bibfield  {journal}
  {\bibinfo  {journal} {Phys. Rev. Lett.}\ }\textbf {\bibinfo {volume} {113}},\
  \bibinfo {pages} {026401} (\bibinfo {year} {2014})}\BibitemShut {NoStop}%
\bibitem [{\citenamefont {Beaud}\ \emph {et~al.}(2014)\citenamefont {Beaud},
  \citenamefont {Caviezel}, \citenamefont {Mariager}, \citenamefont {Rettig},
  \citenamefont {Ingold}, \citenamefont {Dornes}, \citenamefont {Huang},
  \citenamefont {Johnson}, \citenamefont {Radovic}, \citenamefont {Huber},
  \citenamefont {Kubacka}, \citenamefont {Ferrer}, \citenamefont {Lemke},
  \citenamefont {Chollet}, \citenamefont {Zhu}, \citenamefont {Glownia},
  \citenamefont {Sikorski}, \citenamefont {Robert}, \citenamefont {Wadati},
  \citenamefont {Nakamura}, \citenamefont {Kawasaki}, \citenamefont {Tokura},
  \citenamefont {Johnson},\ and\ \citenamefont {Staub}}]{Beaud2014}%
  \BibitemOpen
  \bibfield  {author} {\bibinfo {author} {\bibfnamefont {P.}~\bibnamefont
  {Beaud}}, \bibinfo {author} {\bibfnamefont {A.}~\bibnamefont {Caviezel}},
  \bibinfo {author} {\bibfnamefont {S.~O.}\ \bibnamefont {Mariager}}, \bibinfo
  {author} {\bibfnamefont {L.}~\bibnamefont {Rettig}}, \bibinfo {author}
  {\bibfnamefont {G.}~\bibnamefont {Ingold}}, \bibinfo {author} {\bibfnamefont
  {C.}~\bibnamefont {Dornes}}, \bibinfo {author} {\bibfnamefont {S.-W.}\
  \bibnamefont {Huang}}, \bibinfo {author} {\bibfnamefont {J.~A.}\ \bibnamefont
  {Johnson}}, \bibinfo {author} {\bibfnamefont {M.}~\bibnamefont {Radovic}},
  \bibinfo {author} {\bibfnamefont {T.}~\bibnamefont {Huber}}, \bibinfo
  {author} {\bibfnamefont {T.}~\bibnamefont {Kubacka}}, \bibinfo {author}
  {\bibfnamefont {A.}~\bibnamefont {Ferrer}}, \bibinfo {author} {\bibfnamefont
  {H.~T.}\ \bibnamefont {Lemke}}, \bibinfo {author} {\bibfnamefont
  {M.}~\bibnamefont {Chollet}}, \bibinfo {author} {\bibfnamefont
  {D.}~\bibnamefont {Zhu}}, \bibinfo {author} {\bibfnamefont {J.~M.}\
  \bibnamefont {Glownia}}, \bibinfo {author} {\bibfnamefont {M.}~\bibnamefont
  {Sikorski}}, \bibinfo {author} {\bibfnamefont {A.}~\bibnamefont {Robert}},
  \bibinfo {author} {\bibfnamefont {H.}~\bibnamefont {Wadati}}, \bibinfo
  {author} {\bibfnamefont {M.}~\bibnamefont {Nakamura}}, \bibinfo {author}
  {\bibfnamefont {M.}~\bibnamefont {Kawasaki}}, \bibinfo {author}
  {\bibfnamefont {Y.}~\bibnamefont {Tokura}}, \bibinfo {author} {\bibfnamefont
  {S.~L.}\ \bibnamefont {Johnson}},\ and\ \bibinfo {author} {\bibfnamefont
  {U.}~\bibnamefont {Staub}},\ }\bibfield  {title} {\bibinfo {title} {A
  time-dependent order parameter for ultrafast photoinduced phase
  transitions},\ }\href {https://doi.org/10.1038/nmat4046} {\bibfield
  {journal} {\bibinfo  {journal} {Nat. Mater.}\ }\textbf {\bibinfo {volume}
  {13}},\ \bibinfo {pages} {923} (\bibinfo {year} {2014})}\BibitemShut
  {NoStop}%
\bibitem [{\citenamefont {Teitelbaum}\ \emph {et~al.}(2018)\citenamefont
  {Teitelbaum}, \citenamefont {Shin}, \citenamefont {Wolfson}, \citenamefont
  {Cheng}, \citenamefont {Molesky}, \citenamefont {Kandyla},\ and\
  \citenamefont {Nelson}}]{Teitelbaum2018}%
  \BibitemOpen
  \bibfield  {author} {\bibinfo {author} {\bibfnamefont {S.~W.}\ \bibnamefont
  {Teitelbaum}}, \bibinfo {author} {\bibfnamefont {T.}~\bibnamefont {Shin}},
  \bibinfo {author} {\bibfnamefont {J.~W.}\ \bibnamefont {Wolfson}}, \bibinfo
  {author} {\bibfnamefont {Y.-H.}\ \bibnamefont {Cheng}}, \bibinfo {author}
  {\bibfnamefont {I.~J.~P.}\ \bibnamefont {Molesky}}, \bibinfo {author}
  {\bibfnamefont {M.}~\bibnamefont {Kandyla}},\ and\ \bibinfo {author}
  {\bibfnamefont {K.~A.}\ \bibnamefont {Nelson}},\ }\bibfield  {title}
  {\bibinfo {title} {Real-time observation of a coherent lattice transformation
  into a high-symmetry phase},\ }\href
  {https://doi.org/10.1103/PhysRevX.8.031081} {\bibfield  {journal} {\bibinfo
  {journal} {Phys. Rev. X}\ }\textbf {\bibinfo {volume} {8}},\ \bibinfo {pages}
  {031081} (\bibinfo {year} {2018})}\BibitemShut {NoStop}%
\bibitem [{\citenamefont {Hu}\ \emph {et~al.}(2023)\citenamefont {Hu},
  \citenamefont {Wu}, \citenamefont {Schnyder},\ and\ \citenamefont
  {Shi}}]{Hu2023}%
  \BibitemOpen
  \bibfield  {author} {\bibinfo {author} {\bibfnamefont {Y.}~\bibnamefont
  {Hu}}, \bibinfo {author} {\bibfnamefont {X.}~\bibnamefont {Wu}}, \bibinfo
  {author} {\bibfnamefont {A.~P.}\ \bibnamefont {Schnyder}},\ and\ \bibinfo
  {author} {\bibfnamefont {M.}~\bibnamefont {Shi}},\ }\bibfield  {title}
  {\bibinfo {title} {Electronic landscape of kagome superconductors {$AV_3Sb_5$
  ( A = K, Rb, Cs)} from angle-resolved photoemission spectroscopy},\ }\href
  {https://doi.org/10.1038/s41535-023-00599-y} {\bibfield  {journal} {\bibinfo
  {journal} {npj Quantum Materials}\ }\textbf {\bibinfo {volume} {8}},\
  \bibinfo {pages} {67} (\bibinfo {year} {2023})}\BibitemShut {NoStop}%
\bibitem [{\citenamefont {Morosan}\ \emph {et~al.}(2006)\citenamefont
  {Morosan}, \citenamefont {Zandbergen}, \citenamefont {Dennis}, \citenamefont
  {Bos}, \citenamefont {Onose}, \citenamefont {Klimczuk}, \citenamefont
  {Ramirez}, \citenamefont {Ong},\ and\ \citenamefont {Cava}}]{Morosan2006}%
  \BibitemOpen
  \bibfield  {author} {\bibinfo {author} {\bibfnamefont {E.}~\bibnamefont
  {Morosan}}, \bibinfo {author} {\bibfnamefont {H.~W.}\ \bibnamefont
  {Zandbergen}}, \bibinfo {author} {\bibfnamefont {B.~S.}\ \bibnamefont
  {Dennis}}, \bibinfo {author} {\bibfnamefont {J.~W.~G.}\ \bibnamefont {Bos}},
  \bibinfo {author} {\bibfnamefont {Y.}~\bibnamefont {Onose}}, \bibinfo
  {author} {\bibfnamefont {T.}~\bibnamefont {Klimczuk}}, \bibinfo {author}
  {\bibfnamefont {A.~P.}\ \bibnamefont {Ramirez}}, \bibinfo {author}
  {\bibfnamefont {N.~P.}\ \bibnamefont {Ong}},\ and\ \bibinfo {author}
  {\bibfnamefont {R.~J.}\ \bibnamefont {Cava}},\ }\bibfield  {title} {\bibinfo
  {title} {Superconductivity in {Cu$_x$TiSe$_2$}},\ }\href
  {https://doi.org/10.1038/nphys360} {\bibfield  {journal} {\bibinfo  {journal}
  {Nat. Phys.}\ }\textbf {\bibinfo {volume} {2}},\ \bibinfo {pages} {544}
  (\bibinfo {year} {2006})}\BibitemShut {NoStop}%
\bibitem [{\citenamefont {Yang}\ \emph {et~al.}(2019)\citenamefont {Yang},
  \citenamefont {Tang}, \citenamefont {Duan}, \citenamefont {Zhou},
  \citenamefont {Hao},\ and\ \citenamefont {Zhang}}]{Yang2019}%
  \BibitemOpen
  \bibfield  {author} {\bibinfo {author} {\bibfnamefont {Y.}~\bibnamefont
  {Yang}}, \bibinfo {author} {\bibfnamefont {T.}~\bibnamefont {Tang}}, \bibinfo
  {author} {\bibfnamefont {S.}~\bibnamefont {Duan}}, \bibinfo {author}
  {\bibfnamefont {C.}~\bibnamefont {Zhou}}, \bibinfo {author} {\bibfnamefont
  {D.}~\bibnamefont {Hao}},\ and\ \bibinfo {author} {\bibfnamefont
  {W.}~\bibnamefont {Zhang}},\ }\bibfield  {title} {\bibinfo {title} {A time-
  and angle-resolved photoemission spectroscopy with probe photon energy up to
  6.7 ev},\ }\href {https://doi.org/10.1063/1.5090439} {\bibfield  {journal}
  {\bibinfo  {journal} {Rev. Sci. Instrum.}\ }\textbf {\bibinfo {volume}
  {90}},\ \bibinfo {pages} {063905} (\bibinfo {year} {2019})}\BibitemShut
  {NoStop}%
\bibitem [{\citenamefont {Huang}\ \emph {et~al.}(2022)\citenamefont {Huang},
  \citenamefont {Duan},\ and\ \citenamefont {Zhang}}]{Huang2022}%
  \BibitemOpen
  \bibfield  {author} {\bibinfo {author} {\bibfnamefont {C.}~\bibnamefont
  {Huang}}, \bibinfo {author} {\bibfnamefont {S.}~\bibnamefont {Duan}},\ and\
  \bibinfo {author} {\bibfnamefont {W.}~\bibnamefont {Zhang}},\ }\bibfield
  {title} {\bibinfo {title} {High-resolution time- and angle-resolved
  photoemission studies on quantum materials},\ }\href
  {https://doi.org/10.1007/s44214-022-00013-x} {\bibfield  {journal} {\bibinfo
  {journal} {Quantum Front.}\ }\textbf {\bibinfo {volume} {1}},\ \bibinfo
  {pages} {15} (\bibinfo {year} {2022})}\BibitemShut {NoStop}%
\bibitem [{\citenamefont {Duan}\ \emph {et~al.}(2022)\citenamefont {Duan},
  \citenamefont {Wang}, \citenamefont {Yang}, \citenamefont {Huang},
  \citenamefont {Gu}, \citenamefont {Liu},\ and\ \citenamefont
  {Zhang}}]{Duan2022a}%
  \BibitemOpen
  \bibfield  {author} {\bibinfo {author} {\bibfnamefont {S.}~\bibnamefont
  {Duan}}, \bibinfo {author} {\bibfnamefont {S.}~\bibnamefont {Wang}}, \bibinfo
  {author} {\bibfnamefont {Y.}~\bibnamefont {Yang}}, \bibinfo {author}
  {\bibfnamefont {C.}~\bibnamefont {Huang}}, \bibinfo {author} {\bibfnamefont
  {L.}~\bibnamefont {Gu}}, \bibinfo {author} {\bibfnamefont {H.}~\bibnamefont
  {Liu}},\ and\ \bibinfo {author} {\bibfnamefont {W.}~\bibnamefont {Zhang}},\
  }\bibfield  {title} {\bibinfo {title} {{A sample-position-autocorrection
  system with precision better than 1 $\mu$m in angle-resolved photoemission
  experiments}},\ }\href {https://doi.org/10.1063/5.0106299} {\bibfield
  {journal} {\bibinfo  {journal} {Rev. Sci. Instrum.}\ }\textbf {\bibinfo
  {volume} {93}},\ \bibinfo {pages} {103905} (\bibinfo {year}
  {2022})}\BibitemShut {NoStop}%
\end{thebibliography}

%

\newpage

\newpage

\begin{figure*}
\centering\includegraphics[width=1\columnwidth,page=1]{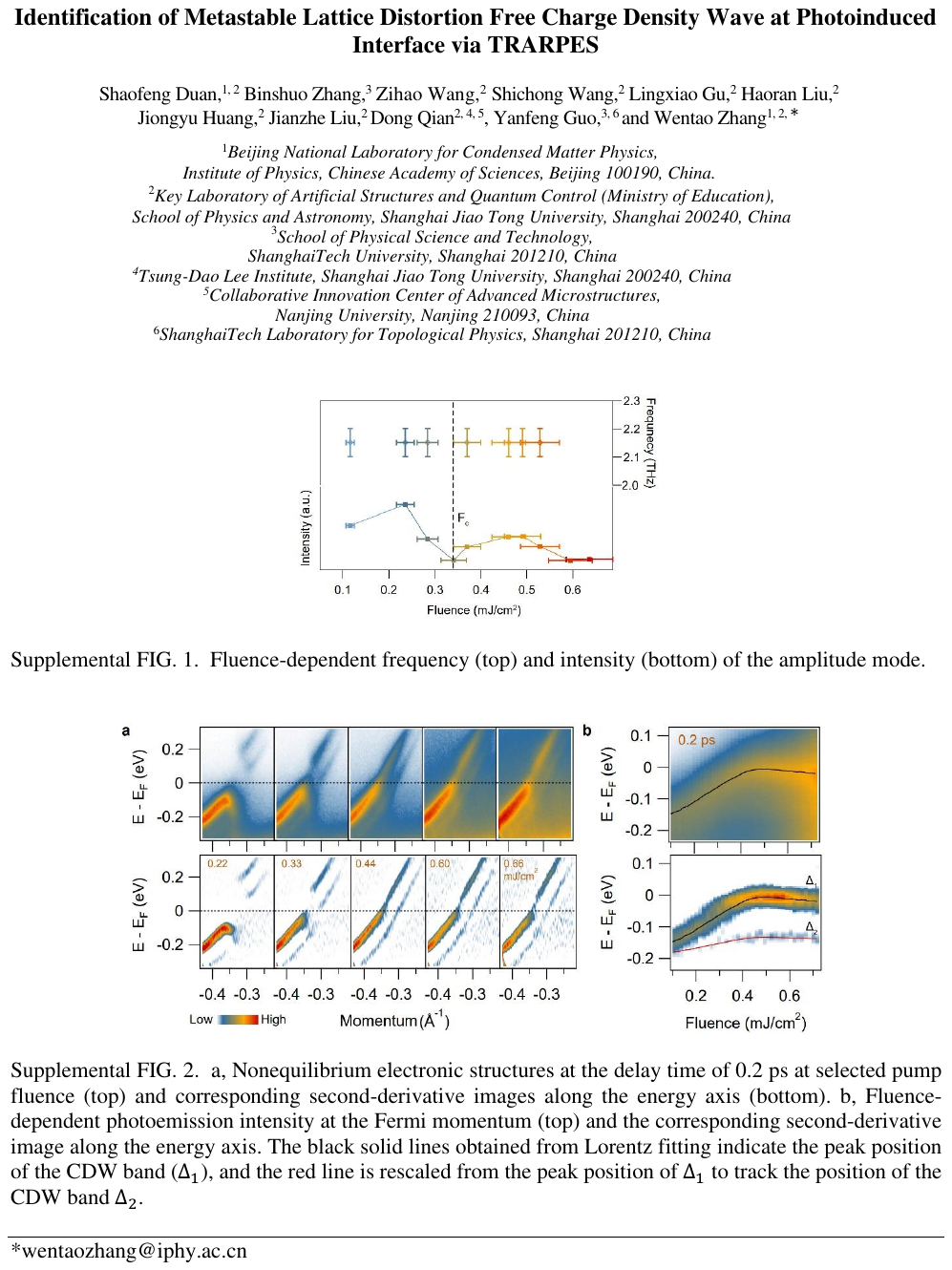}
\end{figure*}
\begin{figure*}
\includegraphics[width=1\columnwidth,page=2]{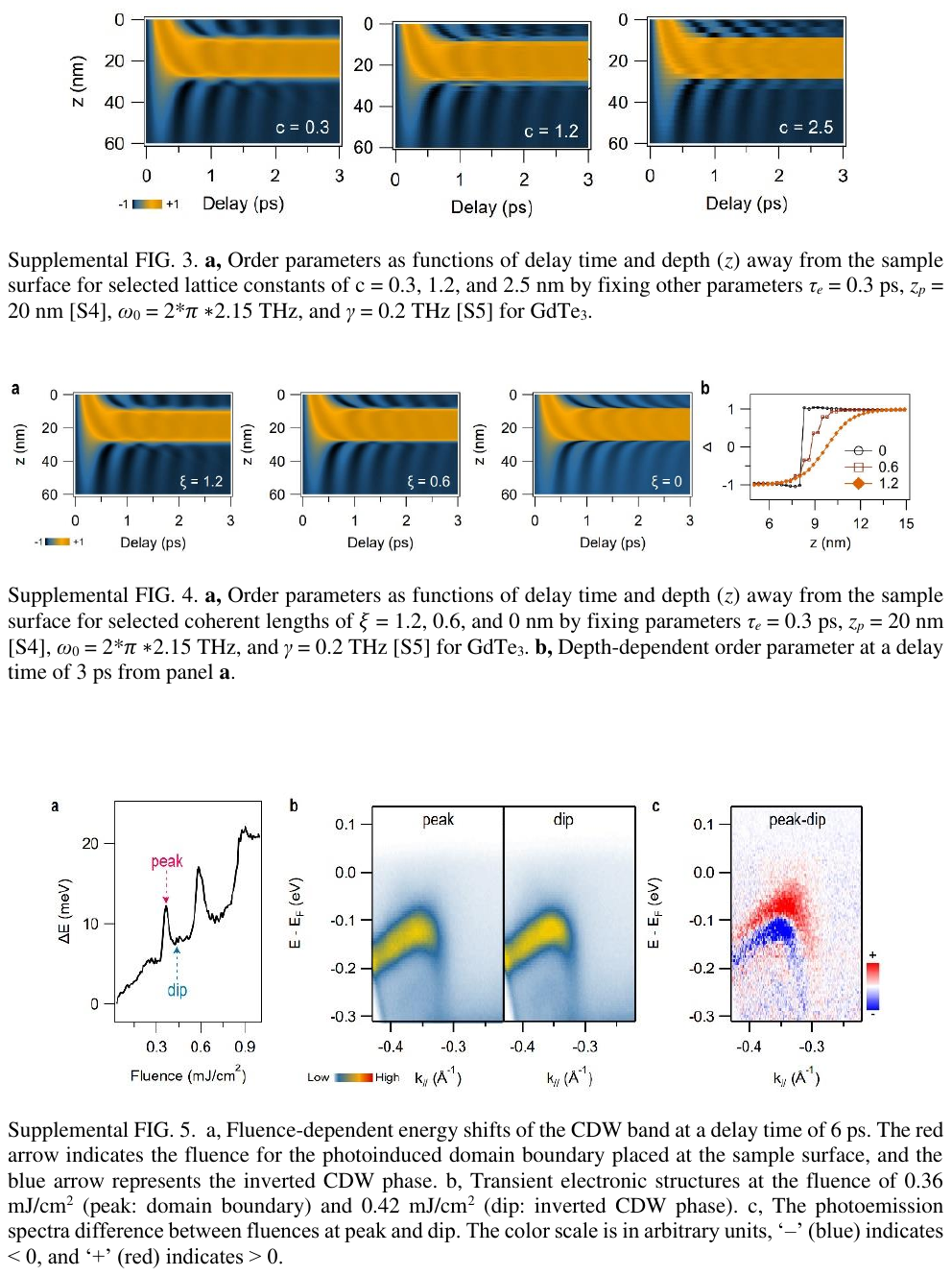}
\end{figure*}
\begin{figure*}
\includegraphics[width=1\columnwidth,page=3]{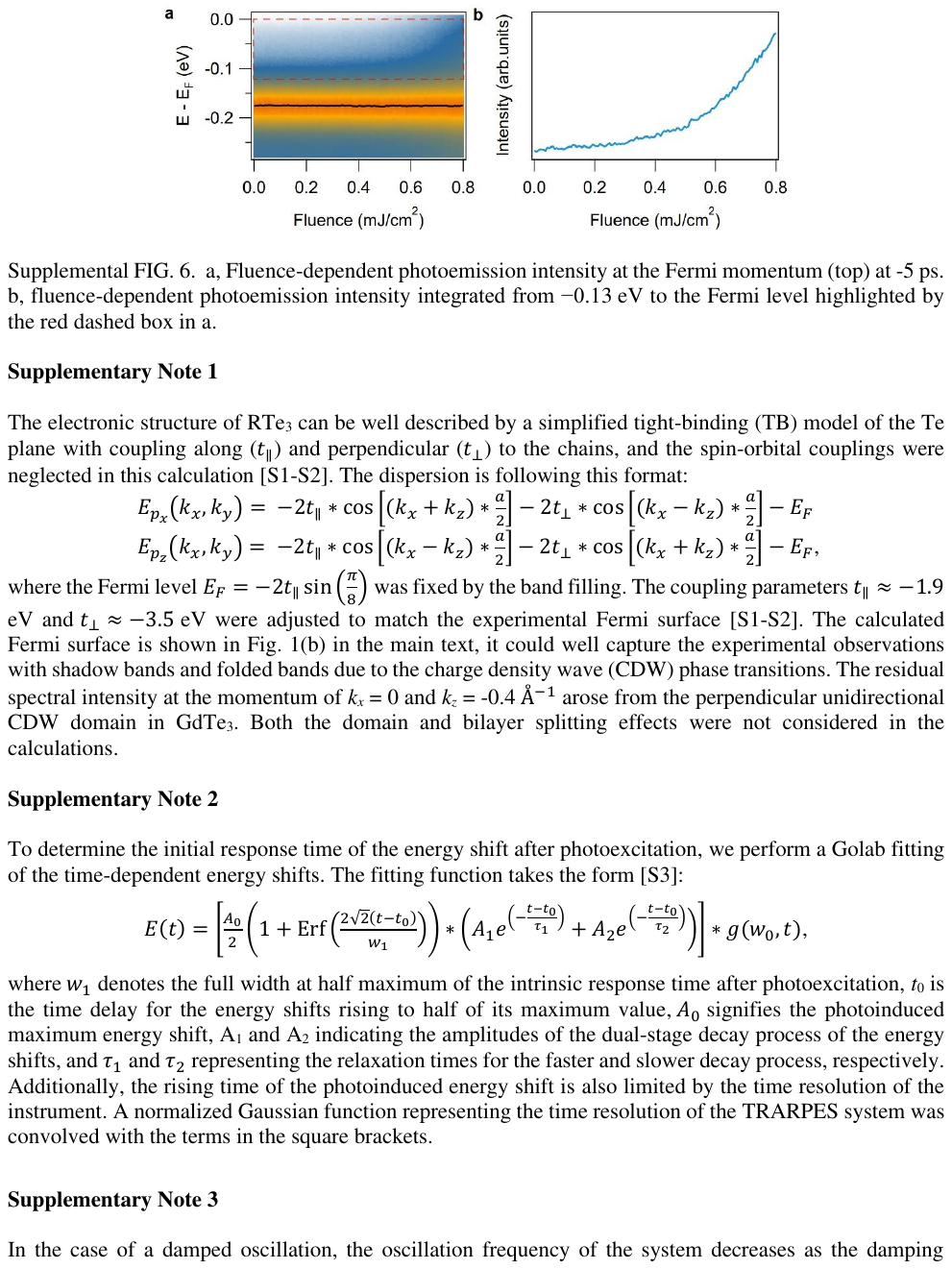}
\end{figure*}
\begin{figure*}
\includegraphics[width=1\columnwidth,page=4]{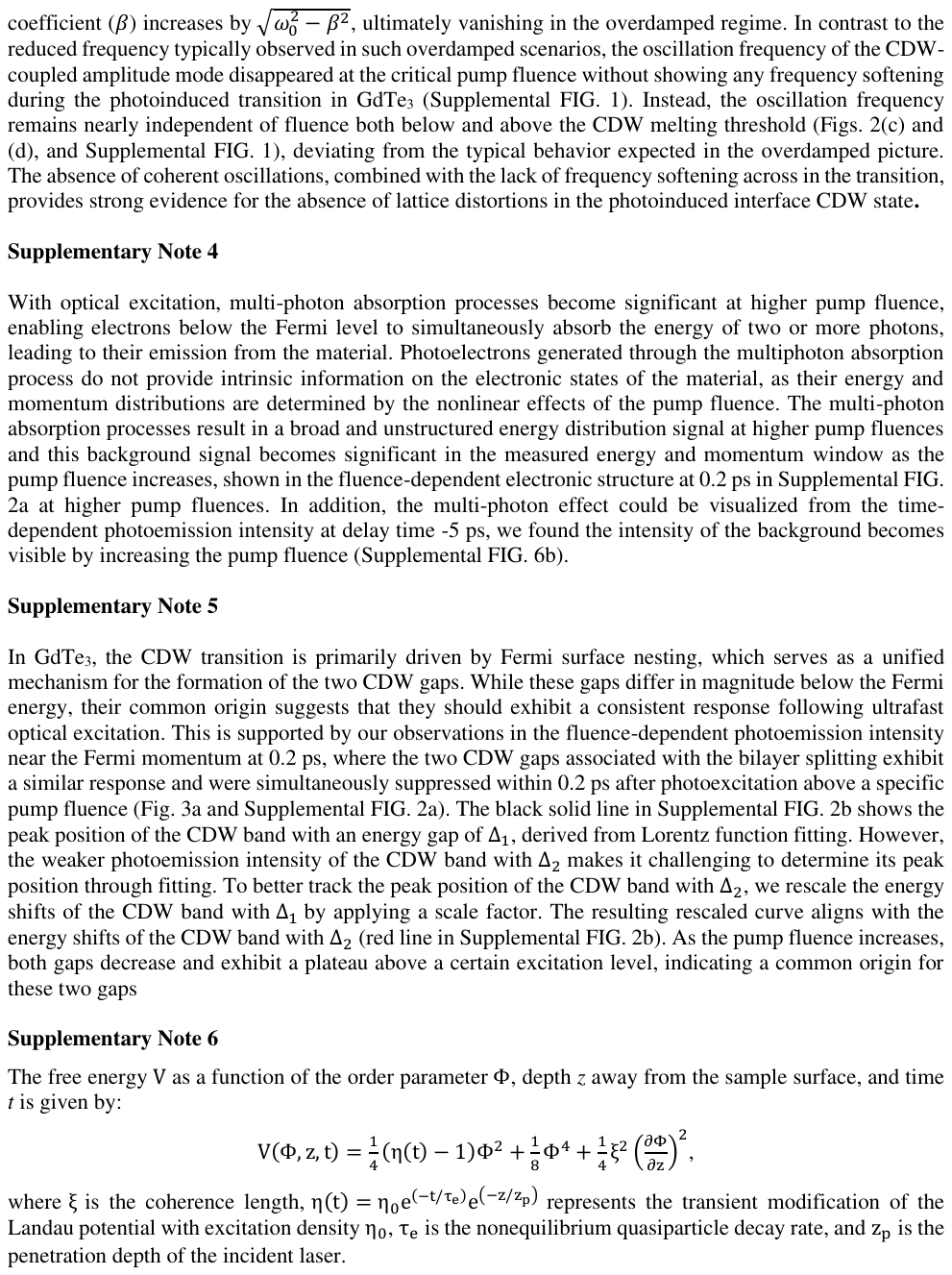}
\end{figure*}
\begin{figure*}
\includegraphics[width=1\columnwidth,page=5]{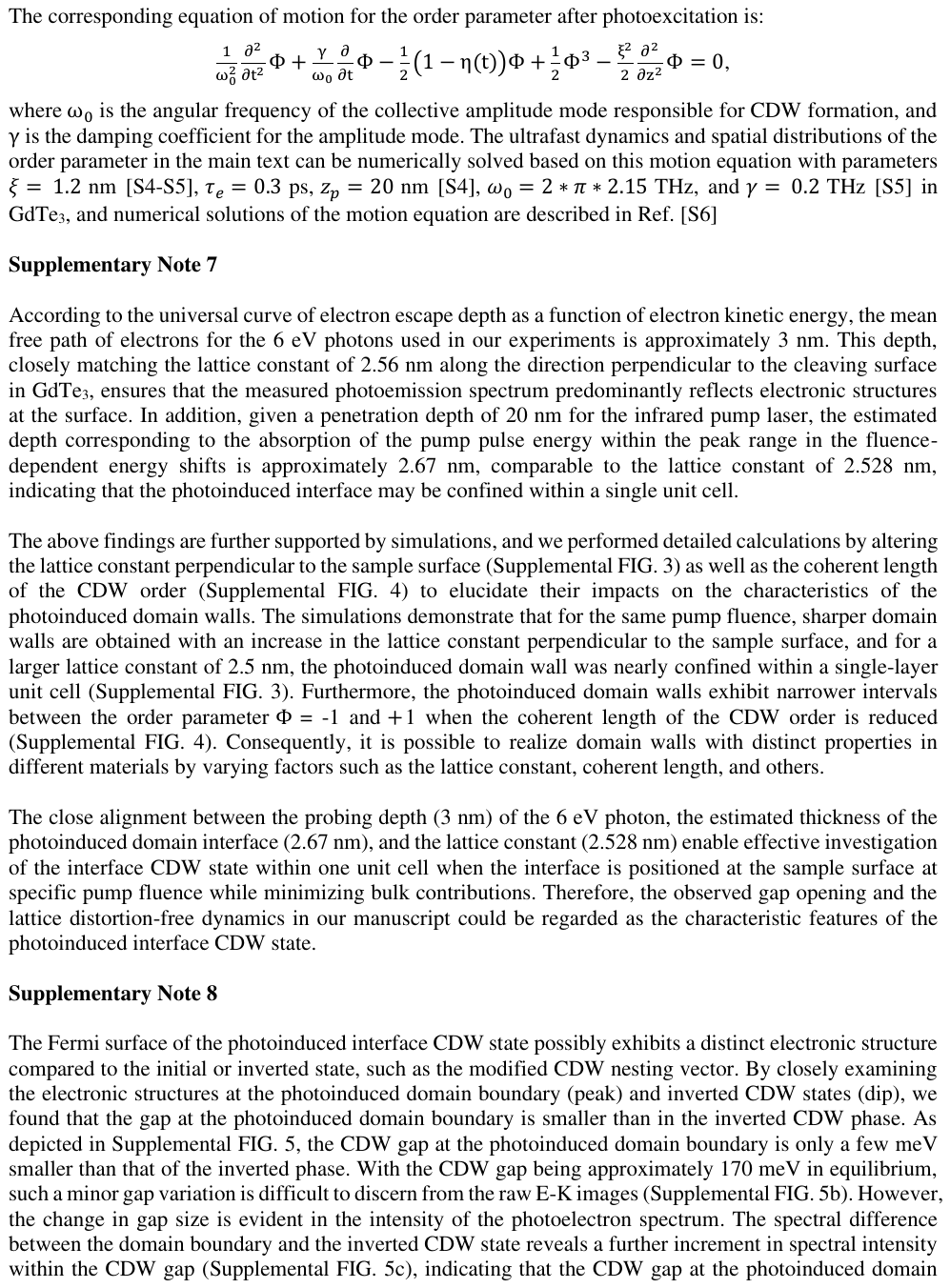}
\end{figure*}
\begin{figure*}
\includegraphics[width=1\columnwidth,page=6]{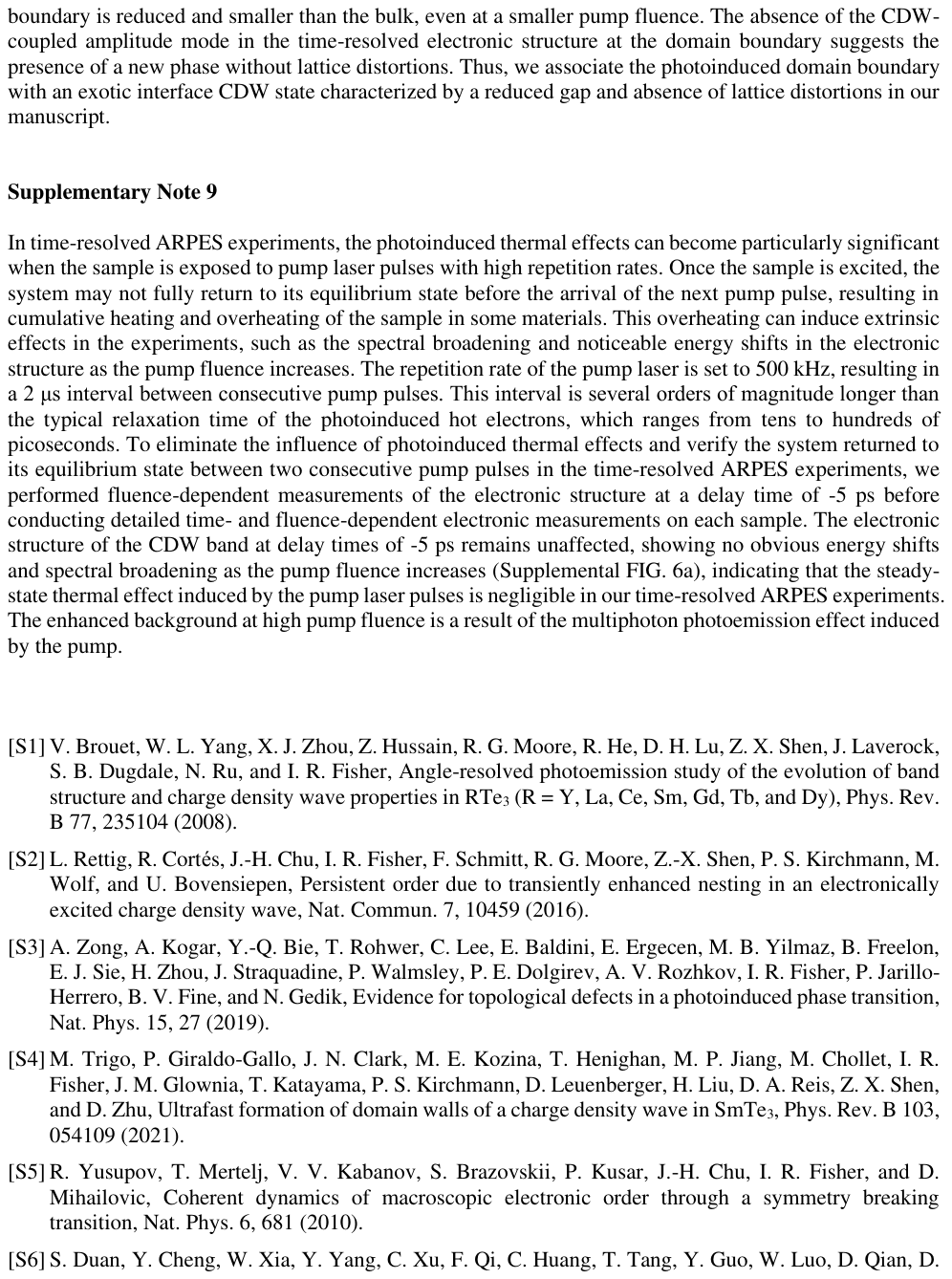}
\end{figure*}

\end{document}